\documentclass[twocolumn]{aastex62}
\usepackage{amssymb,amsmath}
\usepackage{bm}

\submitjournal{The Astrophysical Journal}

\shorttitle{Structure of a protobinary system} 
\shortauthors{Matsumoto et al.}

\begin{document}

\title{
Structure of a Protobinary System: an Asymmetric Circumbinary Disk and Spiral Arms
}

\correspondingauthor{Tomoaki Matsumoto}
\email{matsu@hosei.ac.jp}

\author[0000-0002-8125-4509]{Tomoaki Matsumoto}
\affil{Faculty of Sustainability Studies, Hosei University, Fujimi, Chiyoda-ku, Tokyo 102-8160, Japan}
\affil{Department of Astrophysical Sciences, Princeton University, 4 Ivy Lane, Princeton, NJ 08544, USA}
\affil{RIKEN Center for Computational Science (R-CCS), 7-1-26 Minatojima-minami-machi, Chuo-ku, Kobe, Hyogo 650-0047, Japan}

\author{Kazuya Saigo}
\affil{Chile Observatory, National Astronomical Observatory of Japan, Osawa 2-21-1, Mitaka, Tokyo 181-8588, Japan}

\author[0000-0003-0845-128X]{Shigehisa Takakuwa}
\affil{Department of Physics and Astronomy, Graduate School of Science and Engineering, Kagoshima University, 1-21-35 Korimoto, Kagoshima 890-0065, Japan}
\affil{Academia Sinica Institute of Astronomy and Astrophysics, No. 1, Section 4, Roosevelt Road, Taipei 10617, Taiwan}

\begin{abstract}
We investigate the gas structures around young binary stars by using three-dimensional numerical simulations. Each model exhibits circumstellar disks, spiral arms, and a circumbinary disk with an inner gap or cavity. The circumbinary disk has an asymmetric pattern rotating at an angular velocity of approximately one-fourth of the binary orbit of the moderate-temperature models. Because of this asymmetry, the circumbinary disk has a density bump and a vortex, both of which continue to exist until the end of our calculation. The density bump and vortex are attributed to enhanced angular momentum, which is promoted by the gravitational torque of the stars. In a hot model ($c \ge 2.0$), the asymmetry rotates considerably more slowly than in the moderate-temperature models. The cold models ($c \le 0.02$) exhibit eccentric circumbinary disks, the precession of which is approximated by a secular motion of the ballistic particles. The asymmetry in the circumbinary disk does not depend on the mass ratio, but it becomes less clear as the specific angular momentum of the infalling envelope increases. 
The relative accretion rate onto the stars is sensitive to the angular momentum of the infalling envelope. For envelopes with constant angular momentum, the secondary tends to have a higher accretion rate than the primary, except in very low angular momentum cases. For envelopes with a constant angular velocity, the primary has a higher accretion rate than the secondary because gas with low specific angular momentum falls along the polar directions.
\end{abstract}


\section{Introduction}

More than half of low-mass young stars are members of multiple systems, and multiple star formation is considered to be a major mode of low-mass star formation \citep{Reipurth14}. Recent high-resolution observations taken with the Atacama Large Millimeter/Submillimeter Array (ALMA) have revealed the early phases of low-mass binary and multiple star formation \citep[e.g.,][]{Tokuda14,Dutrey14,Takakuwa14,Tobin16}. In order to compare these high-resolution observations with numerical simulations, the numerical simulations should also be performed at high resolution.

\citet{Takakuwa14,Takakuwa17} observed the Class I binary L1551~NE with ALMA and revealed its detailed circumbinary structures: circumstellar disks, a circumbinary disk with an inner gap or cavity, and spiral arms. They also reported asymmetry in the circumbinary disk with an $m=1$ mode. These characteristic features were reproduced by numerical simulations and synthetic observations. These studies demonstrated the advantages of comparison between observations and simulations, and of modeling a circumbinary system using high-resolution numerical simulations.

To date, there have been many studies on circumbinary systems for young binaries using two-dimensional (2D) smoothed particle hydrodynamics (SPH) numerical simulations \citep{Artymowicz96,Young15a,Dunhill15,Young15b,Nelson16}, 2D grid-based simulations \citep{Gunther02,Ochi05,Hanawa10,deValBorro11,Thun17}, three-dimensional (3D) SPH simulations \citep{Bate97,Satsuka17,Ragusa17,Price18}, and 3D grid-based simulations \citep{Fateeva11,Shi12,Shi15}. There have also been studies on disks surrounding black hole binaries \citep[i.e.,][] {Gomez13,DOrazio13,Farris14,Tang17,Miranda17}.

Although 2D simulations have the advantage of obtaining high resolutions based on the assumption of a thin disk approximation, circumbinary disks are not geometrically thin, as shown in this paper, and 3D simulations are more appropriate for reproducing such disks. 

Several preceding 3D simulations \citep{Shi12,Shi15,Ragusa17} have shown the presence of asymmetries in circumbinary disks, but the physical origin of such asymmetries remains unknown. Moreover, \citet{Shi12} and \citet{Shi15} set the gap of a circumbinary disk outside of the computational domains and did not calculate the flow inside the gap. Such flow would circulate near the inner edge of the gap, as suggested by the 2D simulations \citep[e.g.,][]{Young15a}, where the asymmetry is prominent \citep[e.g.,][]{Ragusa17}.

In this paper, we investigate the circumbinary structures of young binary systems by using 3D numerical simulations with fixed mesh refinement, which allows us to achieve high resolution around the binary stars. The flow inside the gap is also followed at high-resolution. In order to mimic gas accretion onto the binary, we consider infalling envelopes similar to \citet{Bate97} and reproduce the circumstellar structures of the binary systems, including the asymmetries found in circumbinary disks. 

The rest of this paper is organized as follows. In Section~\ref{sec:models} and \ref{sec:methods}, the model and methods are shown. The results of the simulations are presented in Section~\ref{sec:results} and discussed in Section~\ref{sec:discussion}. Finally, the conclusions of this paper are given in Section~\ref{sec:summary}.

\section{Models}
\label{sec:models}

In the 3D computational domain of $[-12D, 12D]^2\times [-6D, 6D]$, two protostars are considered, the barycenter of which is located at the origin, where $D$ is the binary separation. In the initial conditions, the primary and secondary stars are located at $(x,y,z) = (M_2 D /M_\mathrm{tot}, 0, 0)$ and $(x,y,z) = (-M_1 D /M_\mathrm{tot}, 0, 0)$, respectively, where $M_1$ and $M_2$ are masses of the primary and secondary stars, and $M_\mathrm{tot} = M_1 + M_2$ is the total mass of the stars. We assume that the stars rotate around the origin in circular orbits. Since anticlockwise rotation is assumed with an angular velocity of $\Omega_\star = (GM_\mathrm{tot}/D^3)^{1/2}$, the rotation period is $T_\star = 2 \pi (D^3/GM_\mathrm{tot})^{1/2}$, where $G$ is the gravitational constant.

The computational domain is filled with a low density gas of $10^{-3}\rho_0$ at the initial stage. The gas, which has a constant density of $\rho_0$ and a specific angular momentum $j_\mathrm{inf}$, is injected at the cylindrical boundaries: the cylindrical surface of $R=12D$, and the top and bottom surfaces of $z = \pm 6D$, where $R$ denotes the cylindrical radius. On these surfaces, the gas has a radial velocity of $v_r = [2G M_\mathrm{tot}/r - (j_\mathrm{inf}/R)^2]^{1/2}$, assuming freefall from a distance of infinity, where $r$ is the spherical radius. When the centrifugal force is larger than the gravitational force ($j_\mathrm{inf}^2/ R^3 > G M_\mathrm{tot} R /r^3 $), no gas is injected in the area corresponding to the centrifugal barrier. These boundary conditions are similar to \citet{Bate97}. The centrifugal radius is estimated to be $R_\mathrm{cent} = j_\mathrm{inf}^2/(G M_\mathrm{tot})$. An isothermal gas with a constant sound speed of $c$ is assumed.

Two additional model types are considered by changing the boundary condition in Section~\ref{sec:effect_of_infalling_envelope}. In the first type, the gas injection at the boundaries stops ($v_r = 0$) in the course of the calculation. This model examines the effects of the infalling gas. 
In the second type, the gas has a constant angular velocity of $\Omega_\mathrm{inf} = j_\mathrm{inf}/(12D)^2$ and a density distribution of $\rho(r) = \rho_0 (r/12D)^{-2}$ at the boundary surfaces. This boundary condition mimics a rigidly rotating envelope and cloud core, which have often been considered in protostellar collapse simulations \citep[e.g.,][]{Matsumoto03,Machida10}. The gas is, therefore, injected with a specific angular momentum distribution, $j = j_\mathrm{inf} R^2/ (12D)^2$, where $R$ denotes the cylindrical radius of the gas injection points. The gas has a range of specific angular momenta $0 \le j \le j_\mathrm{inf}$ because $R \le 12D$, reproducing the situation where gases with both high and low specific angular momenta fall toward the binary. Two models were considered with $j_\mathrm{inf} = 1.2$ and 2.0. We call these rigidly rotating envelope models.

We ignore the self-gravity of the gas, indicating that the models can be applied to situations in which the total mass of the gas is much less than those of the stars. Therefore, the models are not applicable to binaries just after natal cloud core fragmentation. We also ignore the magnetic field. The effects of these simplifications are discussed in Section~\ref{sec:discussion_modellimit}.

Hereafter, the units of $D=1$, $G M_\mathrm{tot} =1$, and $\rho_0 = 1$ are adopted. The model parameters are the mass ratio of the binary $q = M_2/M_1$, the specific angular momentum of the injected gas $j_\mathrm{inf}$, and the gas sound speed $c$.  The gas temperature is parameterized by $c$.  We construct 23 models by changing $q$, $j_\mathrm{inf}$, and $c$, as shown in Table~\ref{table:model-parameters}.
The model parameters $c$ and $j_\mathrm{inf}$ are shown in units of $\left[(GM_\mathrm{tot})/D\right]^{1/2}$ and $\left(GM_\mathrm{tot}D\right)^{1/2}$, respectively. The normalization of the physical variables is summarized in Table~\ref{table:units}.

\begin{figure*}
\epsscale{1.1}
\plotone{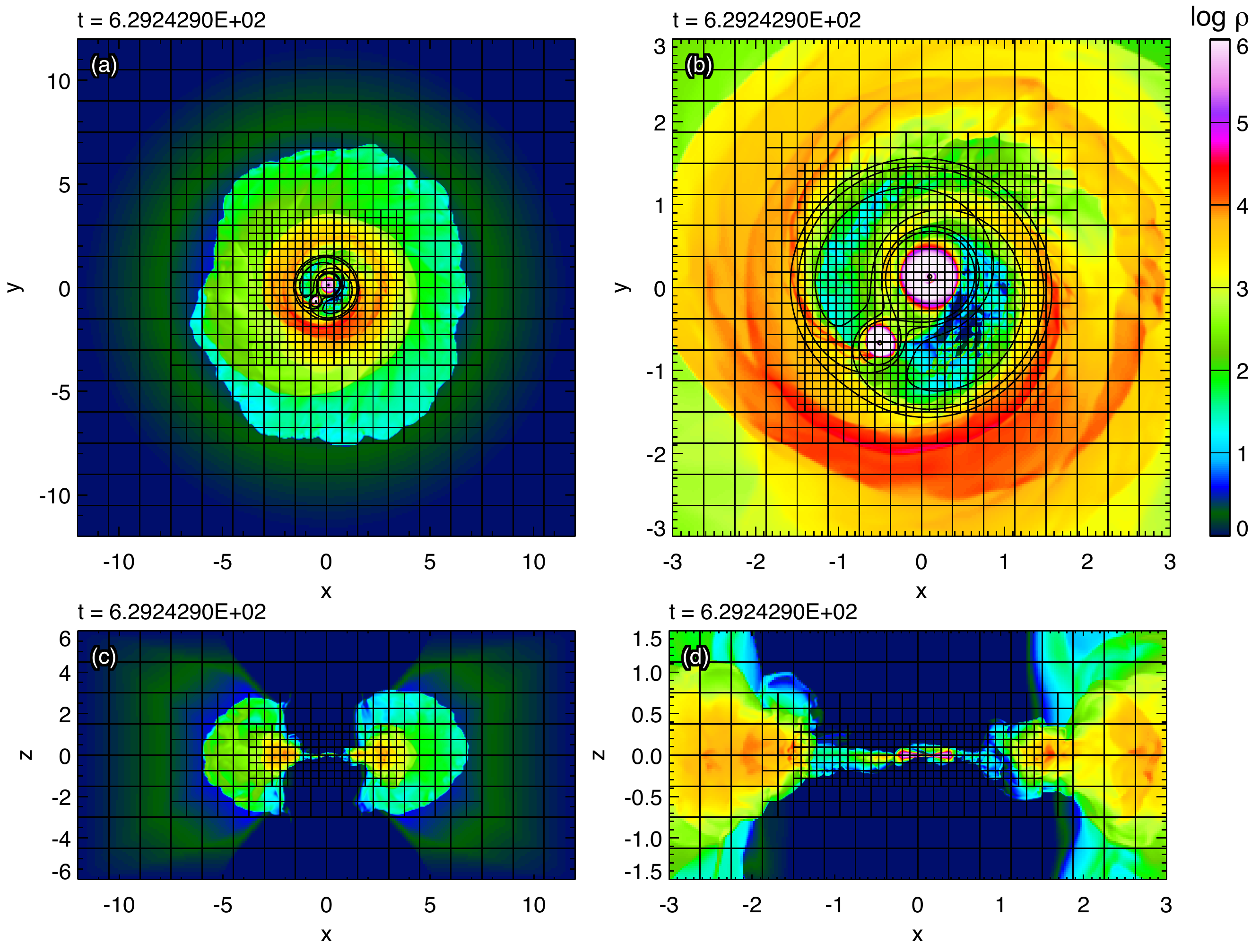}
\caption{
A snapshot of the simulation for the fiducial model with $(q, j_\mathrm{inf}, c) = (0.2, 1.2, 0.1)$ at  $t=629.24$ ($=100.15$ revolutions). The upper panels show the density distributions on a logarithmic scale in the $z=0$ plane, while the lower panels show those in the $y=0$ plane. The right panels are enlargements of the left panels. The squares with the black horizontal and vertical lines depict the AMR blocks. In the upper panels, the contours of the Roche potential are shown for comparison. The contour levels are the potential levels for the L1, L2, and L3 Lagrange points. The coordinates are normalized by the binary separation $D$.
\label{show_amr_grid.pdf}
}
\end{figure*}

\section{Methods}
\label{sec:methods}

The SFUMATO 3D adaptive mesh refinement (AMR) code \citep{Matsumoto07} was employed. The hydrodynamical scheme has third-order accuracy in space with  MUSCL and second-order accuracy in time with the predictor-corrector method. The numerical flux was obtained via a Roe type scheme \citep{Roe81}. The hierarchical grid was fixed during the calculation, and a so-called fixed mesh refinement was then adopted. The grid configuration is shown in Figure~\ref{show_amr_grid.pdf}. The computational domain is divided into $16^2\times8$ blocks for a base grid of $l=0$, and each block has $16^3$ cubic cells. As a result, the base grid has $256^2 \times 128$ cells. The cell width is $\Delta x_\mathrm{min} = 5.86\times10^{-3}$ for the finest grid of $l = 4$, and $\Delta x_\mathrm{max} = 9.38\times10^{-2}$ for the base grid of $l =0$ in units of binary separation.

The binary protostars are represented by two sink particles. The two types of accretion method are implemented in SFUMATO: accretion with a threshold density \citep{Machida10} and Bondi-Hoyle type accretion \citep{Bondi52,Krumholz04}. We adopted the latter type of accretion method here. The detailed implementation of the sink particles is given in \citet{Matsumoto15a}.
The sink particles accrete gas within a sink radius of $r_\mathrm{sink} = 4 \Delta x_\mathrm{min} = 2.34 \times 10^{-2}$, which is considerably smaller than the binary separation.  The masses of the sink particles are fixed in the course of the calculation.

The simulations were performed in the rest frame in this study, although a rotating frame was adopted in \citet{Takakuwa14,Takakuwa17}. In the rest frame, the sink particles rotate in circular orbits in the region of the finest grid ($l = 4$). We confirmed that the difference in the results between the rest and rotating frames was only quantitative. 
The calculation was terminated at 100 revolutions of the binary orbits. Such long-term calculations are required to investigate the flow in a circumbinary system, as indicated by \citet{Young15a}.

\begin{deluxetable}{llllll}
\tablecaption{Model parameters \label{table:model-parameters}}
\tablehead{
\colhead{$q$} &
\colhead{$j_\mathrm{inf}$} &
\colhead{$c$} &
\colhead{$\Omega_p$\tablenotemark{a}} &
\colhead{$\Gamma$\tablenotemark{a}} &
\colhead{Comments}
}
\startdata
0.2  & 1.2 & 0.01  & 0.00504 & 0.229 &\\
0.2  & 1.2 & 0.02  & 0.00676 & 1.13  &\\
0.2  & 1.2 & 0.05  & 0.300   & 3.27  &\\
0.2  & 1.2 & 0.1   & 0.266   & 4.76 &Fiducial model\\
0.2  & 1.2 & 0.133 & 0.249   & 4.89 &\\
0.2  & 1.2 & 0.15  & 0.245   & 4.32 &\\
0.2  & 1.2 & 0.175 & 0.223   & 4.00 &\\
0.2  & 1.2 & 0.2   & 0.0335  & 3.07 &\\
\tableline
0.2  & 0.5 & 0.1 & \nodata   & -1.11 &\\ 
0.2  & 0.8 & 0.1 & \nodata   & 1.60 &\\ 
0.2  & 0.9 & 0.1 & 0.270     & 3.64 &\\
0.2  & 1.0 & 0.1 & 0.266     & 4.25 &\\
0.2  & 1.4 & 0.1 & 0.252     & 5.23 &\\
0.2  & 1.6 & 0.1 & 0.237     & 5.44 &\\
0.2  & 1.8 & 0.1 & \nodata   & 5.39 &\\ 
0.2  & 2.0 & 0.1 & \nodata   & 5.26 &\\ 
\tableline
0.0  & 1.2 & 0.1 & 0.589     & \nodata &\\
0.5  & 1.2 & 0.1 & 0.234     & 1.44 &\\
0.7  & 1.2 & 0.1 & 0.228     & 0.689 &\\
1.0  & 1.2 & 0.1 & 0.225     & 0.0246 &\\
\tableline
0.2  & 1.2 & 0.1 & 0.254     & 2.84 & Gas injection suspended \\ 
0.2  & 1.2 & 0.1 & 0.260     & 1.05 & Rigidly rotating envelope \\ 
0.2  & 2.0 & 0.1 & 0.231     & 0.120 & Rigidly rotating envelope 
\enddata
\tablenotetext{a}{Temporal average in  $t = (160-200) \pi$}
\end{deluxetable}

\begin{deluxetable}{ll}
\tablecaption{Normalization \label{table:units}}
\tablehead{
\colhead{Variables} &
\colhead{Units}
}
\startdata
Length $(x, y, z)$ & $D$\\
Velocity $v$, sound speed $c$   & $\left(GM_\mathrm{tot}/D\right)^{1/2}$ \\
Specific angular momentum $j$ & $\left(GM_\mathrm{tot}D\right)^{1/2}$\\
Time $t$ & $(D^3/GM_\mathrm{tot})^{1/2}$\\
Angular velocity $\Omega$ & $(GM_\mathrm{tot}/D^3)^{1/2}$\\
Density $\rho$ & $\rho_0$\\
Surface density $\Sigma$ & $\rho_0 D$
\enddata
\end{deluxetable}

\section{Results}
\label{sec:results}

\subsection{Overall Evolution of Fiducial Model}
\label{sec:Fiducial_model}

\begin{figure*}
\epsscale{1.1}
\plotone{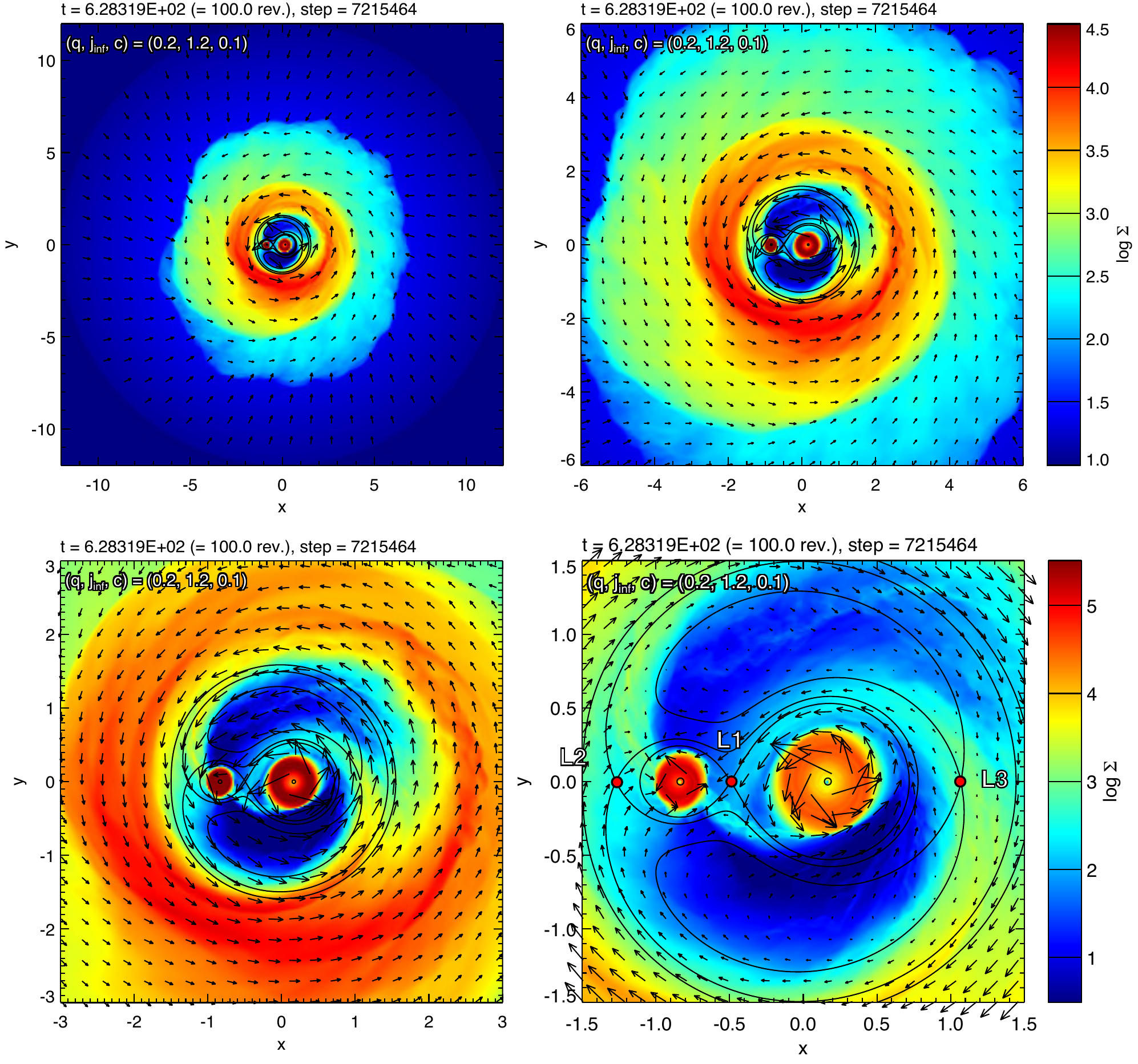}
\caption{
Surface density and velocity distributions on four different scales for the fiducial model with $(q, j_\mathrm{inf}, c) = (0.2, 1.2, 0.1)$ at 100 revolutions of the binary. The color denotes the surface density distribution integrated along the $z$-direction over the computational domain, $\Sigma(x,y) = \int_{-6D}^{6D}\rho(x,y,z) dz$. The top color bar is for the top panels and the bottom-left panel, and the bottom color bar is for the bottom-right panel. The arrows denote the density-weighted velocity distribution in the $x-y$ plane, $(\bar{v}_x, \bar{v}_y) = \int_{-6D}^{6D} (v_x, v_y) \rho dz / \Sigma$. In the lower right panel, the velocity is converted to that seen in the rotating frame with the same angular velocity as the binary, while the other panels display the velocity in the rest frame. The Roche potential, the levels of which are the potential levels of the L1, L2, and L3 Lagrange points, is shown by the contour. The two sink particles are shown by the small circles with a sink radius $r_\mathrm{sink}$. The primary and secondary are located in the right and left sides, respectively. The L2, L1, and L3 Lagrange points lie on the $x$-axis from left to right in each panel.
\label{rbinAcc_p02j12c01v10_arrow.pdf}
 }
\end{figure*}

As a fiducial mode, a model with $(q, j_\mathrm{inf}, c) = (0.2, 1.2, 0.1)$ is examined here because the model parameters are similar to those for the Class I binary L1551~NE \citep{Takakuwa14,Takakuwa17}. The gas is injected at the boundary surfaces and falls toward the stars. At $\sim 5$ revolutions of the binary, a circumbinary disk forms with a radius that is approximately equal to the centrifugal radius of the infalling gas, $R_\mathrm{cent} = 1.44$. Since the primary and secondary stars each have circumstellar disks, they are referred to hereafter as the circumprimary and circumsecondary disks.

Figure~\ref{show_amr_grid.pdf} and \ref{rbinAcc_p02j12c01v10_arrow.pdf} show the volume density and surface density distributions, respectively, at $\sim 100$ revolutions of the binary. At this stage, the circumbinary disk extends up to a radius of $R \sim 6$. The circumbinary disk has a gap with a radius of $R \sim 1.5$, in which both the stars rotate. As shown in Figure~\ref{show_amr_grid.pdf}, the circumstellar disks are geometrically thin, while the circumbinary disk exhibits a thick structure and extends up to $z \sim \pm 3$ in the $z$-direction.

The circumstellar disks have clear outer edges, the sizes of which are consistent with the \citet{Pichardo05} who estimated the disk sizes by exploring test particle orbits. In contrast, the gap radius ($R \sim 1.5$) is smaller than their estimation of $R \sim 1.8$. The gas inflow along the spiral arms may reduce the gap.

Two spiral arms exist in the gap of the circumbinary disk, as shown in Figure~\ref{rbinAcc_p02j12c01v10_arrow.pdf}. Hereafter, we call the spiral arms on the sides of the primary and secondary stars the primary and secondary arms, respectively. The secondary arm is connected to the circumsecondary disk, while the primary arm runs around the circumprimary disk and is connected to the bridge between the circumstellar disks. The spiral arms are caused by the gravitational torque of the stars. The bridge is a shock wave that is caused by the convergence of the gas flows inside the gap. These spiral arms and bridge were commonly seen in previous simulations \citep[e.g.,][]{Bate97}. While \citet{Kaigorodov10} and \citet{Fateeva11} have argued that two bow shocks associated with the circumstellar disks play an important role in opening the gap of the circumbinary disk, no bow shock is associated with the circumsecondary disk (Figure~\ref{rbinAcc_p02j12c01v10_arrow.pdf}). Moreover, the gas flow along the primary arm suggests that the primary arm is a stream rather than a bow shock (see also the streamlines in Figure~\ref{rbinAcc_p02j12c01v10_ua7215464_rotframe.pdf}).

\begin{figure}
 \epsscale{1.1}
 \plotone{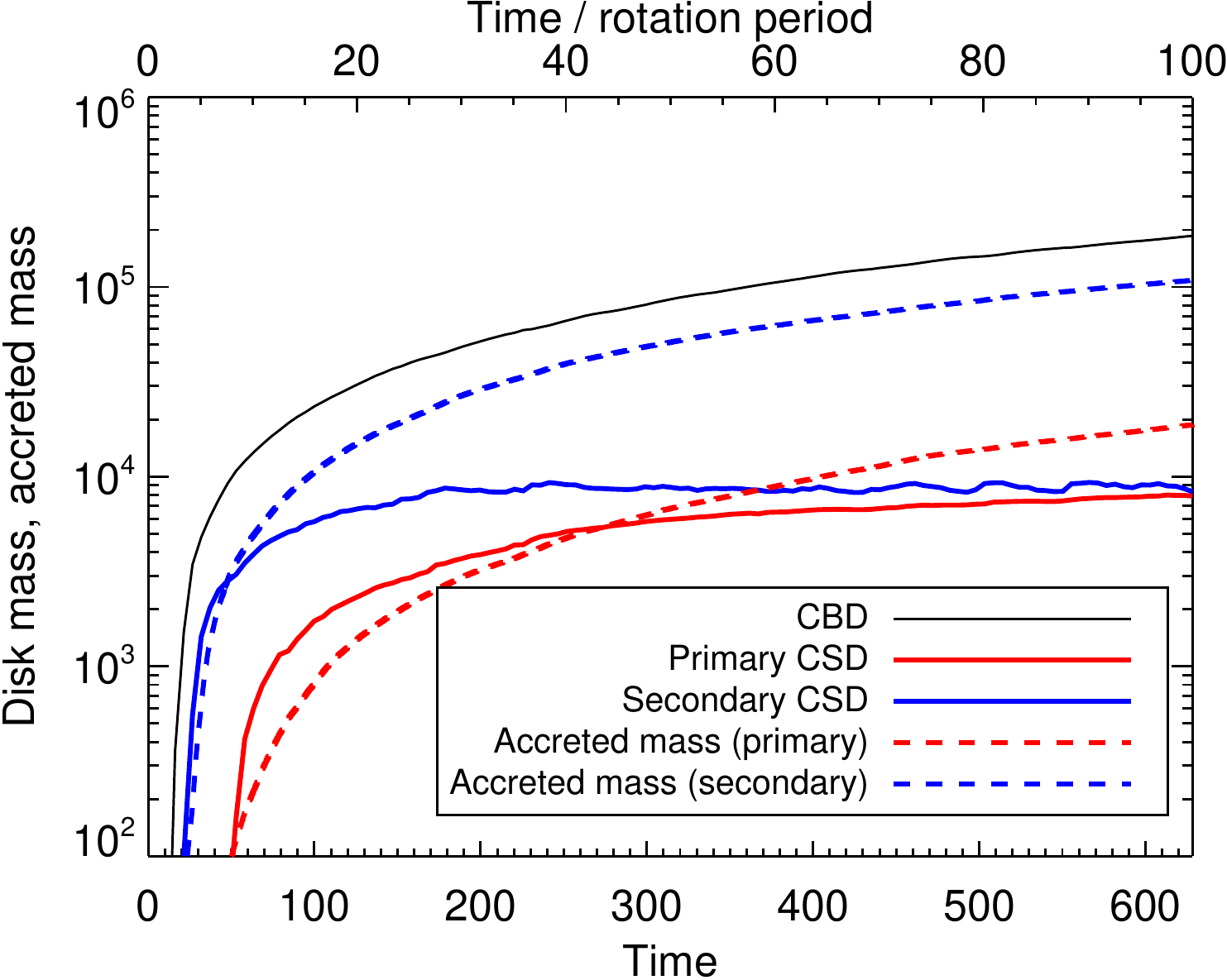}
 \caption{
 Masses of the circumbinary disk (CBD) and the circumstellar disks (CSDs) for the primary and secondary as a function of time for the fiducial model with $(q, j_\mathrm{inf}, c) = (0.2, 1.2, 0.1)$. The dashed lines are the accreted masses to the primary and secondary for comparison. The accreted mass is defined as the accumulated mass of the gas that accretes onto each sink particle.
 \label{disk_total_p02j12c01v10_log.pdf}
 }
\end{figure}

The mass and radius of the circumbinary disk increase as time proceeds (Figure~\ref{disk_total_p02j12c01v10_log.pdf}). The circumsecondary disk reaches a constant mass at  $\sim 30$ revolutions of the binary, while the circumprimary disk continues to show an increasing trend at 100 revolutions. At the last stage, the ratio of the masses of the circumbinary disk, circumprimary disk, and circumsecondary disk is $M_\mathrm{CBD}: M_\mathrm{CPD}: M_\mathrm{CSD} = 1.00: 0.043: 0.045$.

Of the gas injected at the boundary surfaces, $\sim 40\%$ is eventually accreted onto either of the sink particles (see dashed lines in Figure~\ref{disk_total_p02j12c01v10_log.pdf}) and the remaining $\sim 60\%$ accumulates in the circumbinary disk, which grows in mass and radius. Because the growth of the circumbinary disk is slow compared to the rotation timescale, $\dot{M}_\mathrm{CBD}/M_\mathrm{CBD} \sim 10^{-3}$ at $\sim 100$ revolutions, the structure of the circumbinary disk exhibits a quasi-steady state during the simulation. In the measurement of the masses of the disks, each circumstellar disk is defined by gas with a density higher than $10^5$ that is located inside each Roche lobe, and the circumbinary disk is defined by gas with a density higher than 10 that is located outside the two Roche lobes (see Figure~\ref{show_amr_grid.pdf}).


Note that the accreted masses predominate over the masses of the circumstellar disk for both the primary and secondary in the long-term simulation here. This suggests that models without sink particles are suitable for short-term simulations \citep[e.g,][]{Ochi05,Hanawa10}.

\subsection{Gas Flow in the Circumbinary System}

\begin{figure*}
 \epsscale{0.5}
\includegraphics[height=0.4\textwidth]{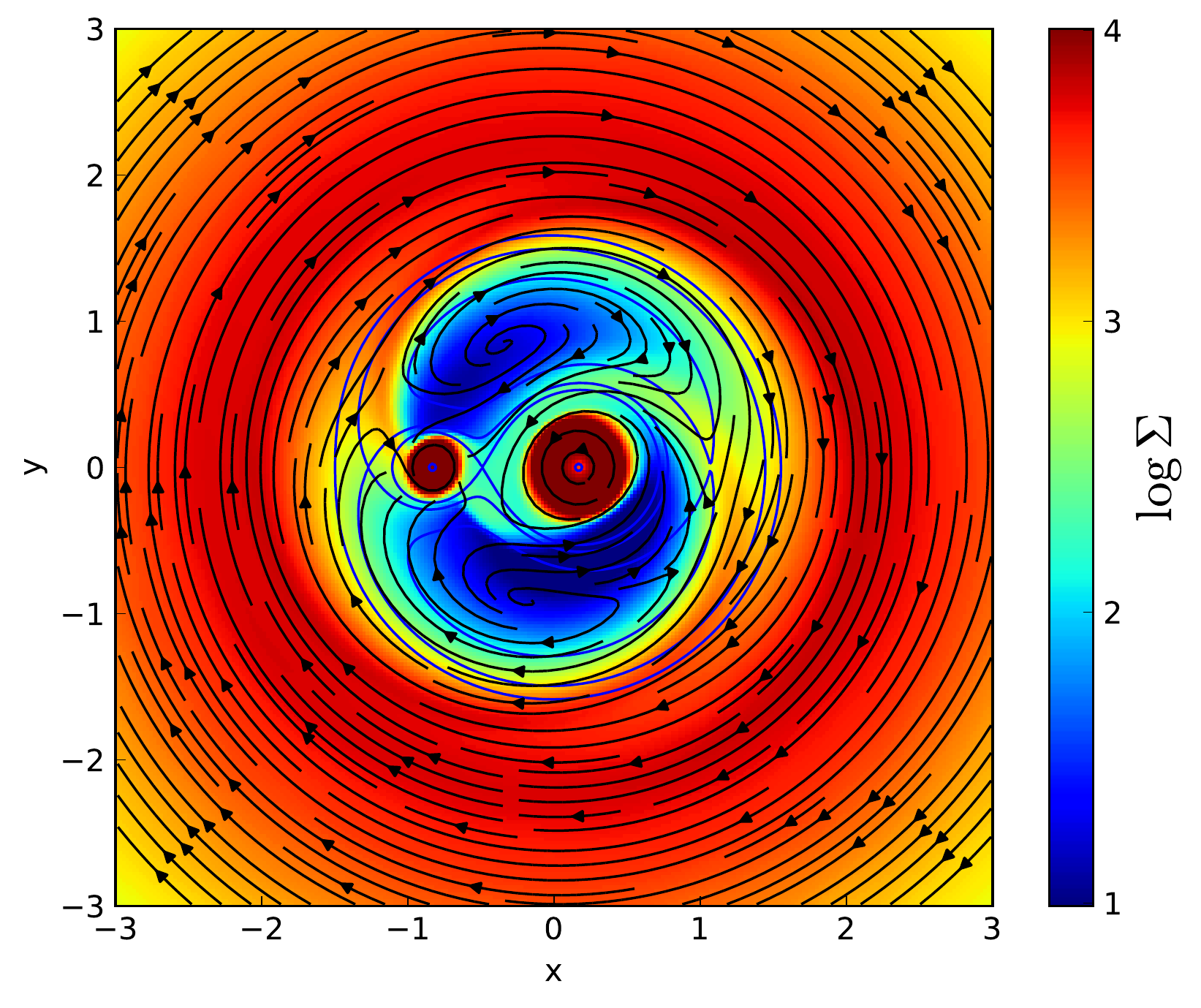}
\includegraphics[height=0.4\textwidth]{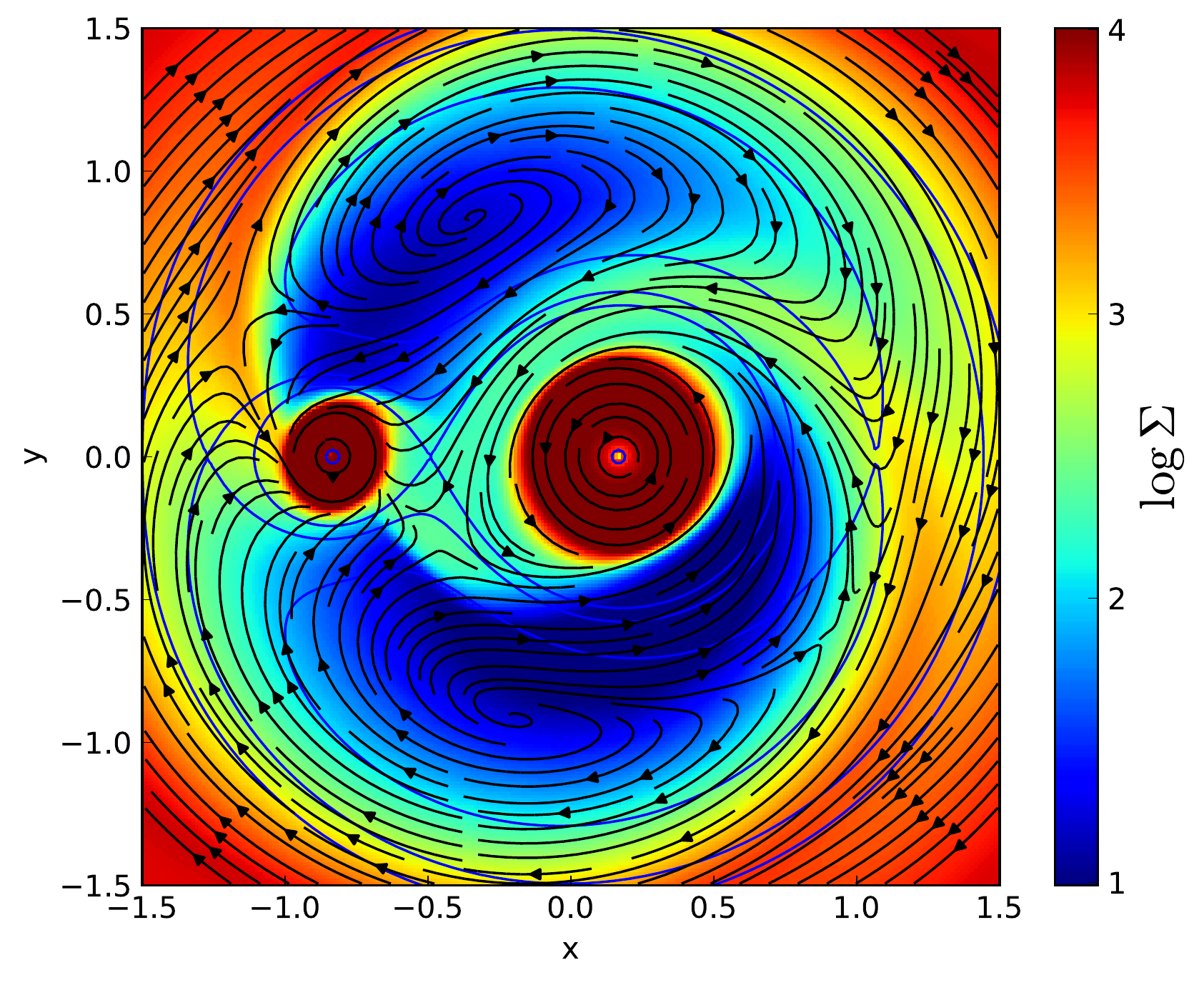}
 \caption{
 Temporal average of surface density and velocity distributions for the fiducial model with $(q, j_\mathrm{inf}, c) = (0.2, 1.2, 0.1)$. The temporal average is taken in the period from 80 to 100 revolutions of the binary, and is also taken in a rotating frame with the same angular velocity as the binary. The color and streamlines denote the surface density and the density-weighted velocity distributions, respectively. The right panel is a magnification of the left panel. The Roche potential and the sink particles are shown by the blue lines.
 \label{rbinAcc_p02j12c01v10_ua7215464_rotframe.pdf}
 }
\end{figure*}

The circumbinary structures exhibit small-scale fluctuations. In order to erase such fluctuations, the temporal average is taken over the period of $80-100$ revolutions, as shown in Figure~\ref{rbinAcc_p02j12c01v10_ua7215464_rotframe.pdf}. 
The purpose of the temporal average is to investigate the dynamics of the gas, not for comparison with observations. With the temporally averaged distributions, we can present long-lived, essential physical phenomena in the circumbinary disk.
While the temporally averaged surface density exhibits a smooth distribution compared to the snapshot, the spiral arms are clearly seen. The bridge between the circumstellar disks is blurred in the temporally averaged surface density because the bridge changes as it wavers in the vicinity of the L1 point. 

The streamlines in Figure~\ref{rbinAcc_p02j12c01v10_ua7215464_rotframe.pdf} show the temporally averaged, density-weighted velocity distribution. The gas falls along the spiral arms from the circumbinary disk to the circumstellar disks. 
The gas stream along the secondary arms falls directly onto the circumsecondary disk. The gas stream along the primary arm is connected to the bridge between the two circumstellar disks. These streams along the spiral arms cause the accretion from the circumbinary disk to the circumstellar disks.

The streamlines show eddy-like streams around the L4 and L5 points in the gap of the circumbinary disk. These streams are similar to a tadpole orbit, which is one of the closed orbits in the restricted three-body problem. The eddy-like streams are smoothly connected to those in the spiral arms and the circumbinary disk.
Note that there are two stagnation points near the L2 and L3 points at $(x,y) \simeq (-1.2, 0.5)$ and $(0.9, -0.6)$, respectively.

\begin{figure*}
 \epsscale{0.36}
 \plotone{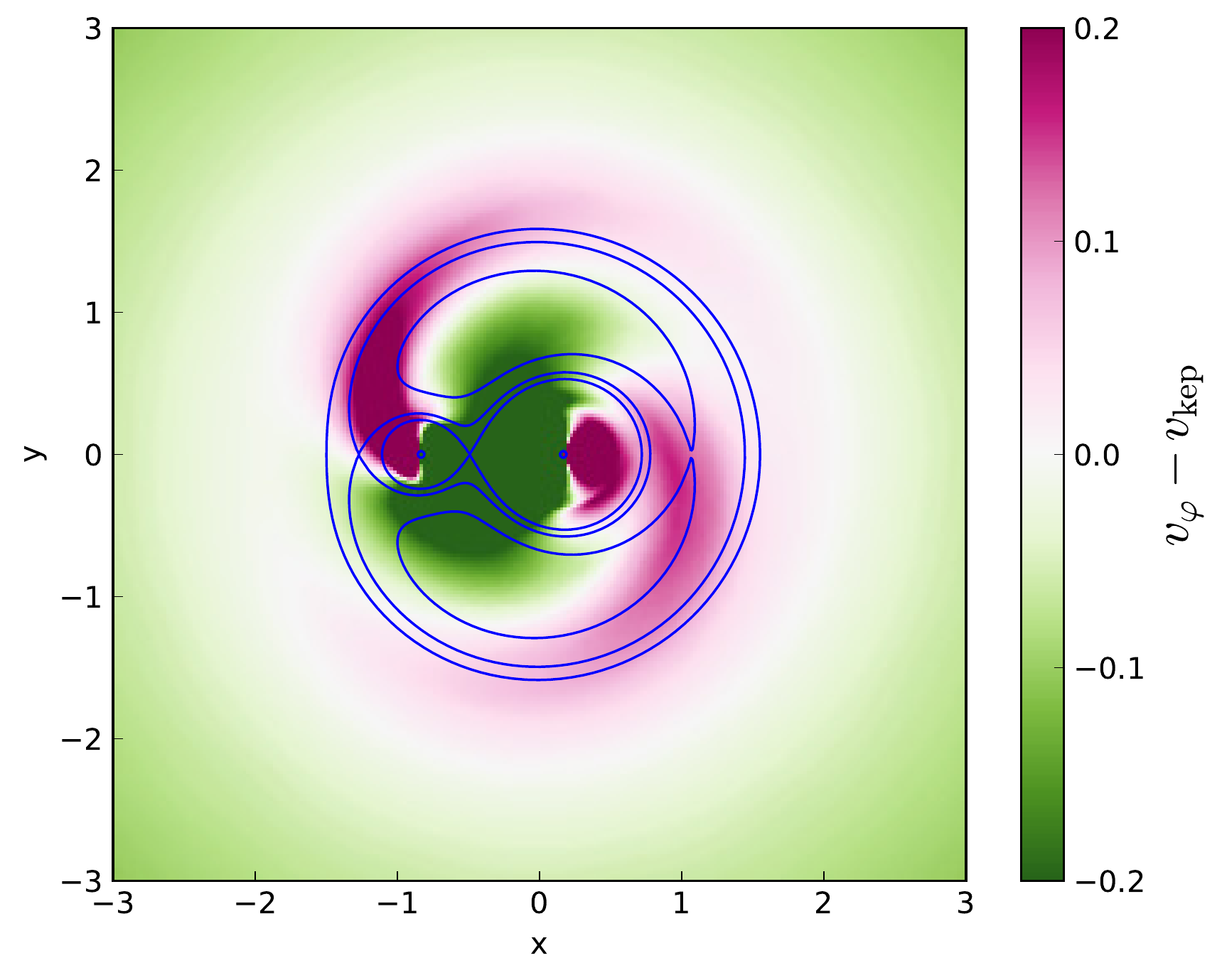}
 \plotone{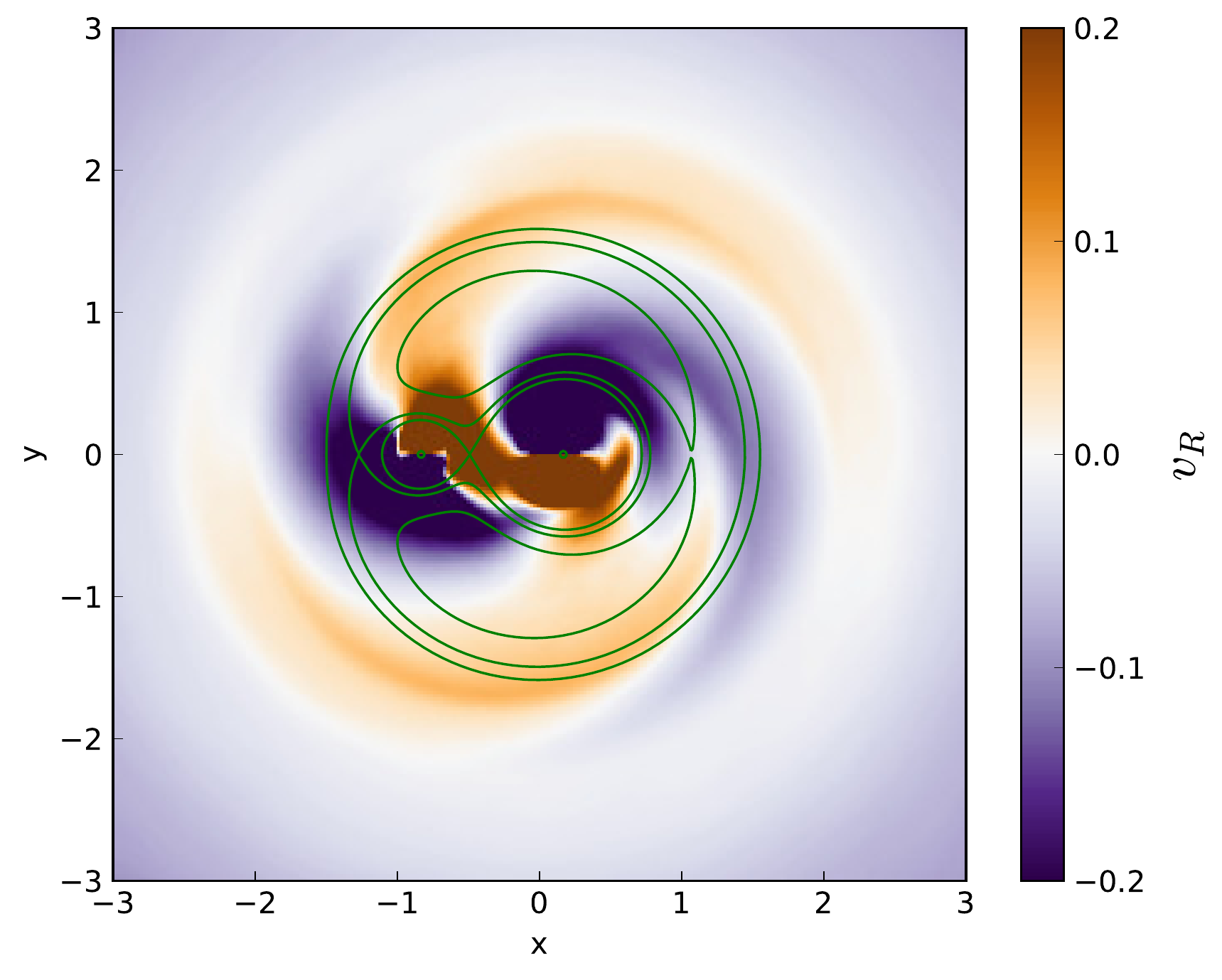}
 \plotone{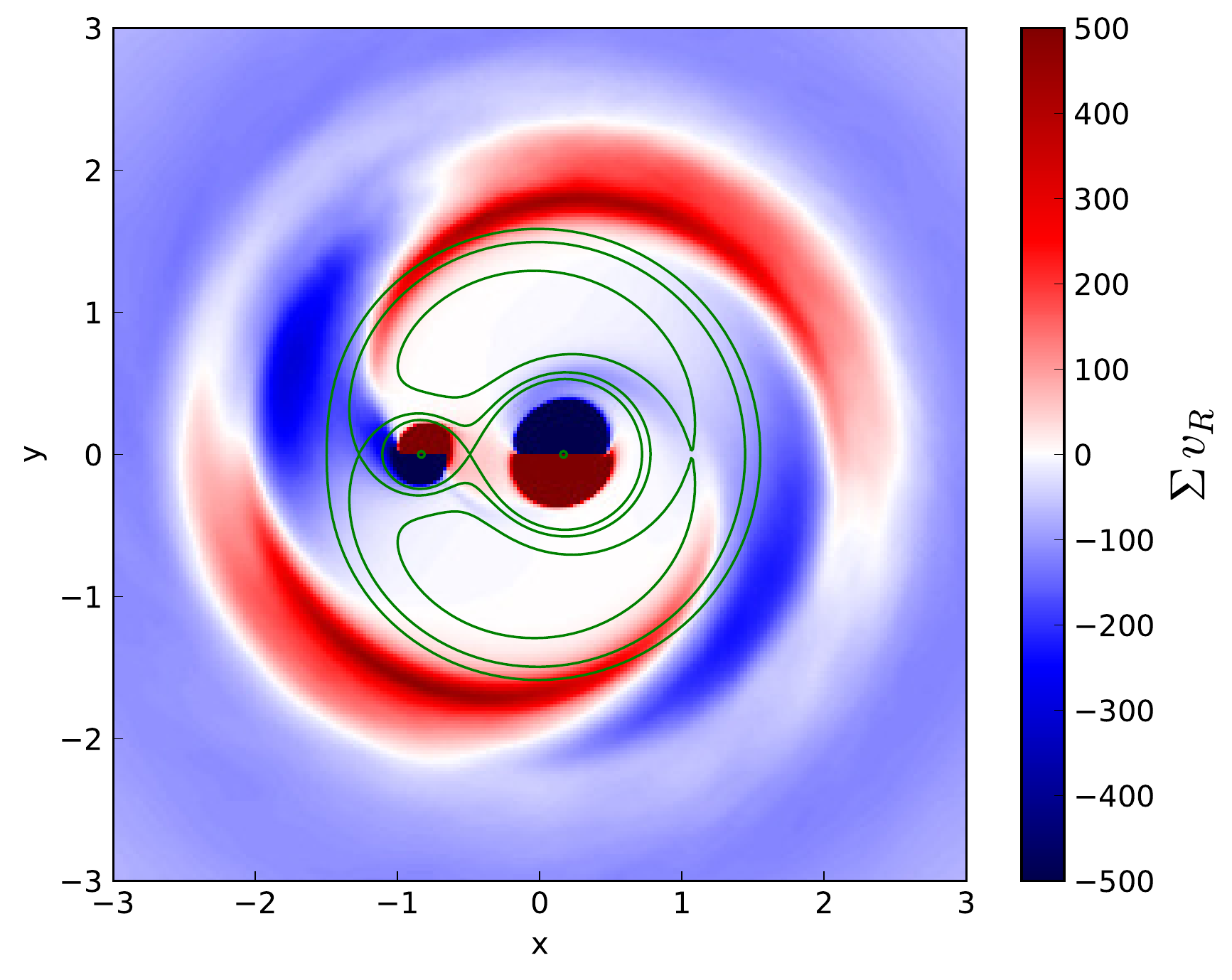}
 \caption{ Temporal averages of the difference between the rotation velocity and the Keplerian velocity $v_\varphi - v_\mathrm{kep}$, where the Keplerian velocity is defined by $v_\mathrm{kep} = (GM_\mathrm{tot}/r)^{1/2}$ (left panel), the radial velocity $v_R$ (middle panel), and the radial mass flux $\Sigma v_R$ (right panel) for the fiducial model with $(q, j_\mathrm{inf}, c) = (0.2, 1.2, 0.1)$. The temporal average is calculated during the period from 80 to 100 revolutions of the binary and in a rotating frame with the same angular velocity as the binary. All the values are density-weighted averages along the $z$-direction. The velocity is measured in the rest frame. The stage and region plotted here are the same as those in Figure~\ref{rbinAcc_p02j12c01v10_ua7215464_rotframe.pdf} (left).
 \label{rbinAcc_p02j12c01v10_ua7215464_vp.pdf}
 }
\end{figure*}

Figure~\ref{rbinAcc_p02j12c01v10_ua7215464_vp.pdf} shows the gas flow more quantitatively. The two spiral arms have a rotation velocity faster than the Keplerian velocity (Figure~\ref{rbinAcc_p02j12c01v10_ua7215464_vp.pdf}, left panel). The gas of the spiral arms is accelerated by the gravitational torque from the orbiting stars, thereby showing fast rotation.

Along the spiral arms, the radial velocity indicates both infall and outflow (Figure~\ref{rbinAcc_p02j12c01v10_ua7215464_vp.pdf}, middle panel). The inner portion of each spiral arm exhibits inflow, while the outer portion exhibits outflow. The boundary between the inflow and outflow coincides with the location of the stagnation point near the L2 or L3 point for each spiral arm. Note that the circumbinary disk (outside the outermost contour of the Roche potential) has both inflow and outflow regions. In the circumbinary disk, outflow regions coincide with the spiral arms while inflow regions coincide with the inter-arm regions. Such inflow and outflow are caused by gravitational torque from the orbiting stars, and have definitely been detected by ALMA observations \citep{Takakuwa14}.

\begin{figure}
 \epsscale{1.1}
 \plotone{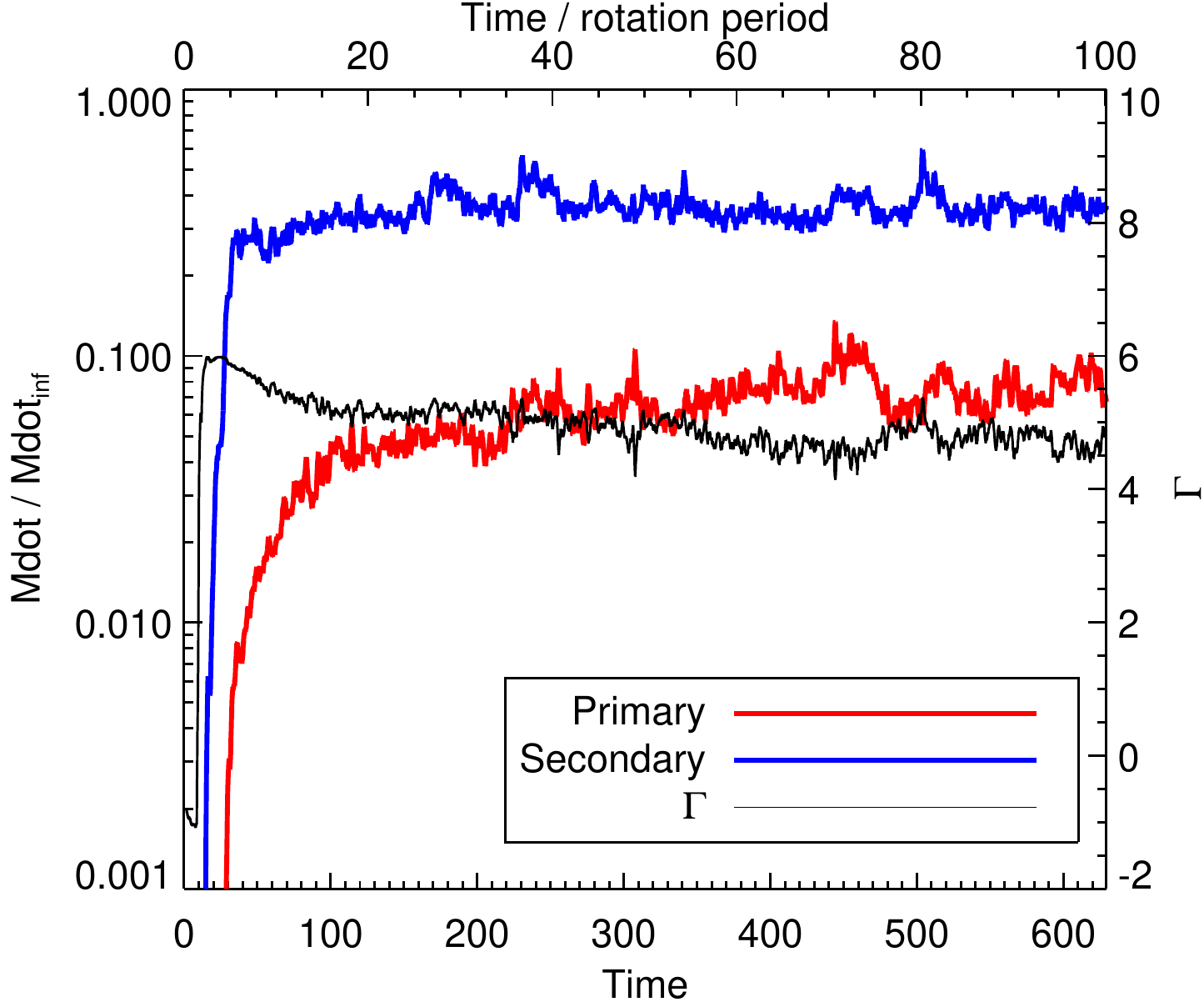}
 \caption{Accretion rates for the primary (red line) and secondary (blue line) as a function of time for the fiducial model with $(q, j_\mathrm{inf}, c) = (0.2, 1.2, 0.1)$. The accretion rates are normalized by the mass injection rate of the boundary condition. The time in the upper abscissa is normalized by the rotation period of the binary. The change rate for the mass ratio $\Gamma$ is also shown (black line).
 \label{plot_pmass.pdf}
 }
\end{figure}

The radial mass flux is examined by using the distribution of $\Sigma v_r$ in the $x-y$ plane (Figure~\ref{rbinAcc_p02j12c01v10_ua7215464_vp.pdf}, right panel). Inside the gap, the secondary arm exhibits higher inflow (deeper blue) than the primary arm does. This is consistent with the fact that the secondary has a higher accretion rate than the primary (Figure~\ref{plot_pmass.pdf}). The bridge between the circumprimary and circumsecondary disks shows a positive radial flux (pale red), indicating that the gas is transferred from the primary side to the secondary side over the bridge. However, this flux is considerably lower than that through the secondary arm, thereby indicating that accretion onto the circumsecondary disk occurs mainly through the secondary arm.

\subsection{Accretion Rates}

Figure~\ref{plot_pmass.pdf} shows the accretion rates for the primary and secondary for the fiducial model. To reduce the short-term variability of the accretion rates, we measured the accretion rates as average rates, $\dot{M}_i(t) = (M_i(t-\Delta t/2) - M_i(t+\Delta t/2))/\Delta t$ for $i = 1,2$ (the primary and secondary), where $M_i$ denotes the accumulated mass of the gas that accretes onto each stars, and $\Delta t \simeq 1.5$ ($\simeq 0.25$ rotation period). The accretion rate for the secondary ($\dot{M}_2$) reaches a constant value by the early phase, while that for the primary ($\dot{M}_1$) gradually increases with significant modulation. The secondary has a higher accretion rate than the primary by a factor of $\dot{M}_2/ \dot{M}_1 \sim 4-6$ after $t \sim 300$. 

Figure~\ref{plot_pmass.pdf} also shows the dimensionless variable of 
\begin{equation}
\Gamma = \frac{\dot{q}/q}{\dot{M}_\mathrm{tot}/M_\mathrm{tot}} = 
\frac{(1+q) \left(\dot{M}_2 - q \dot{M}_1 \right)}{q \left(\dot{M}_1 + \dot{M}_2\right)} \;.
\end{equation}
The variable $\Gamma$ is the change rate of the mass ratio $q$ normalized by the accretion rate onto both of the stars. The fiducial model exhibits $\Gamma \sim 4.5 - 5$, which is in agreement with a corresponding model in \citet{Bate97} and \citet{Young15a}. The temporal average of the last 20 revolutions is $\Gamma = 4.76 $ (Table~\ref{table:model-parameters}). The positive $\Gamma$ indicates an increase in the mass ratio.
 
\subsection{Asymmetry in Circumbinary Disk}
\label{sec:asymmetry_in_cbd}
For many models, a circumbinary disk shows asymmetry in the density and surface density distributions. For the fiducial model with $(q, j_\mathrm{inf}, c) = (0.2, 1.2, 0.1)$, as shown in Figure~\ref{show_amr_grid.pdf}b, the circumbinary disk has higher density on the lower side at the stage shown there. In Figure~\ref{rbinAcc_p02j12c01v10_arrow.pdf}, the surface density of the circumbinary disk is higher on the lower left side. The asymmetric pattern rotates in the azimuthal direction at an angular velocity lower than that of the binary $\Omega_\star$. Therefore, the relative position angle between the asymmetry and the binary stars changes as time proceeds. 

Note that the spiral arms rotate at the binary angular velocity $\Omega_\star$ in the gap and inner region ($R \lesssim 1.5$ for the fiducial model) of the circumbinary disk. The waves of spiral arms accumulate to form a local density bump in the circumbinary disk, which causes the asymmetry there, as can be clearly observed in Figure~\ref{rbinAcc_p02j12c01v10_arrow.pdf}.

\begin{figure}
\epsscale{1.1}
\plotone{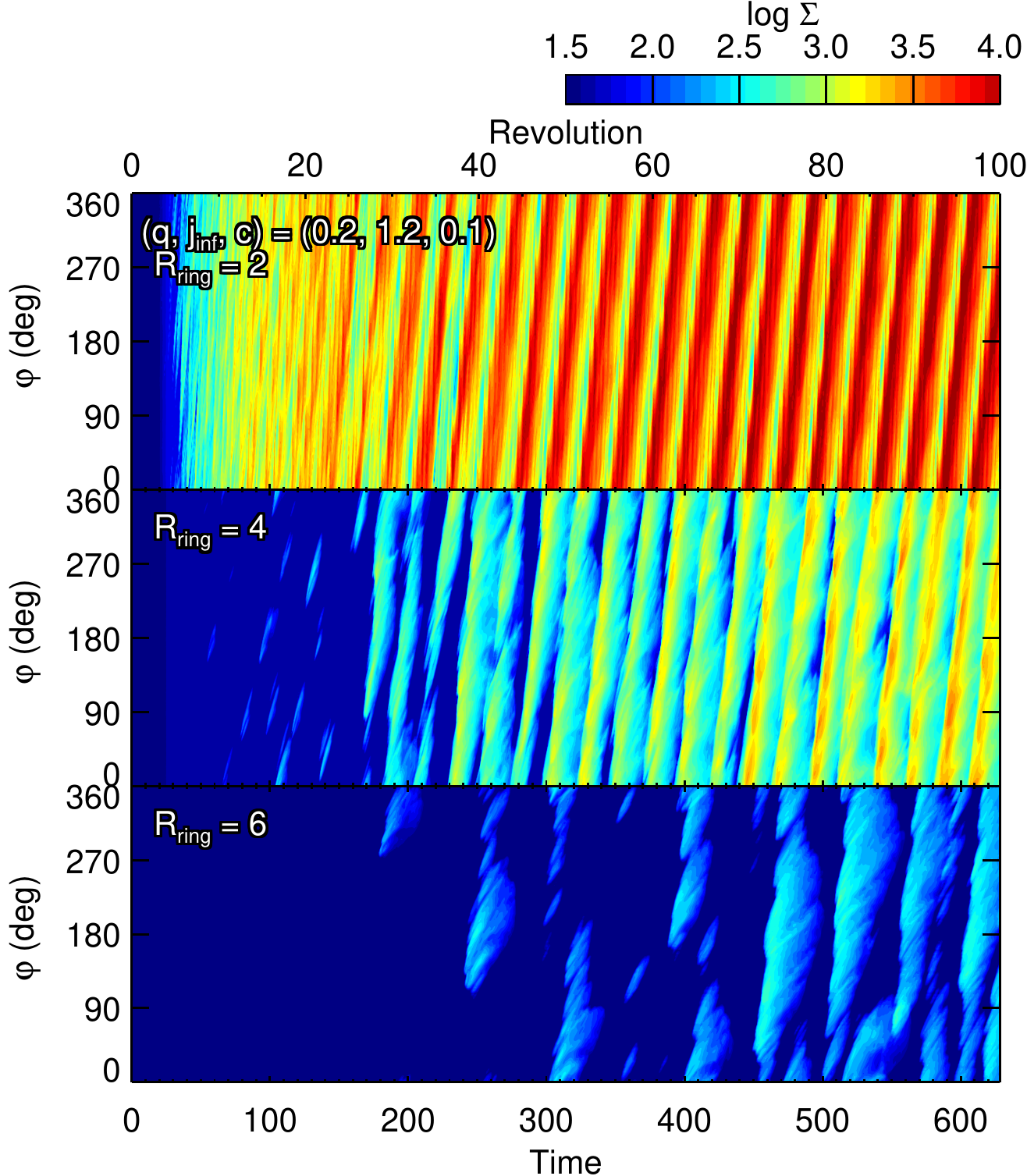}
\caption{
 Surface density distribution in the cylindrical shells as a function of time and azimuthal angle, $\varphi$, for the model with $(q, j_\mathrm{inf}, c) = (0.2, 1.2, 0.1)$. The upper abscissa shows the time in units of the binary rotation period. The radii of the cylindrical shells are $R_\mathrm{ring} = 2$, 4, and 6 from top to bottom.
 \label{disk_oscillate_diagram1.pdf}
 }
\end{figure}

\begin{figure}
\epsscale{1.1}
\plotone{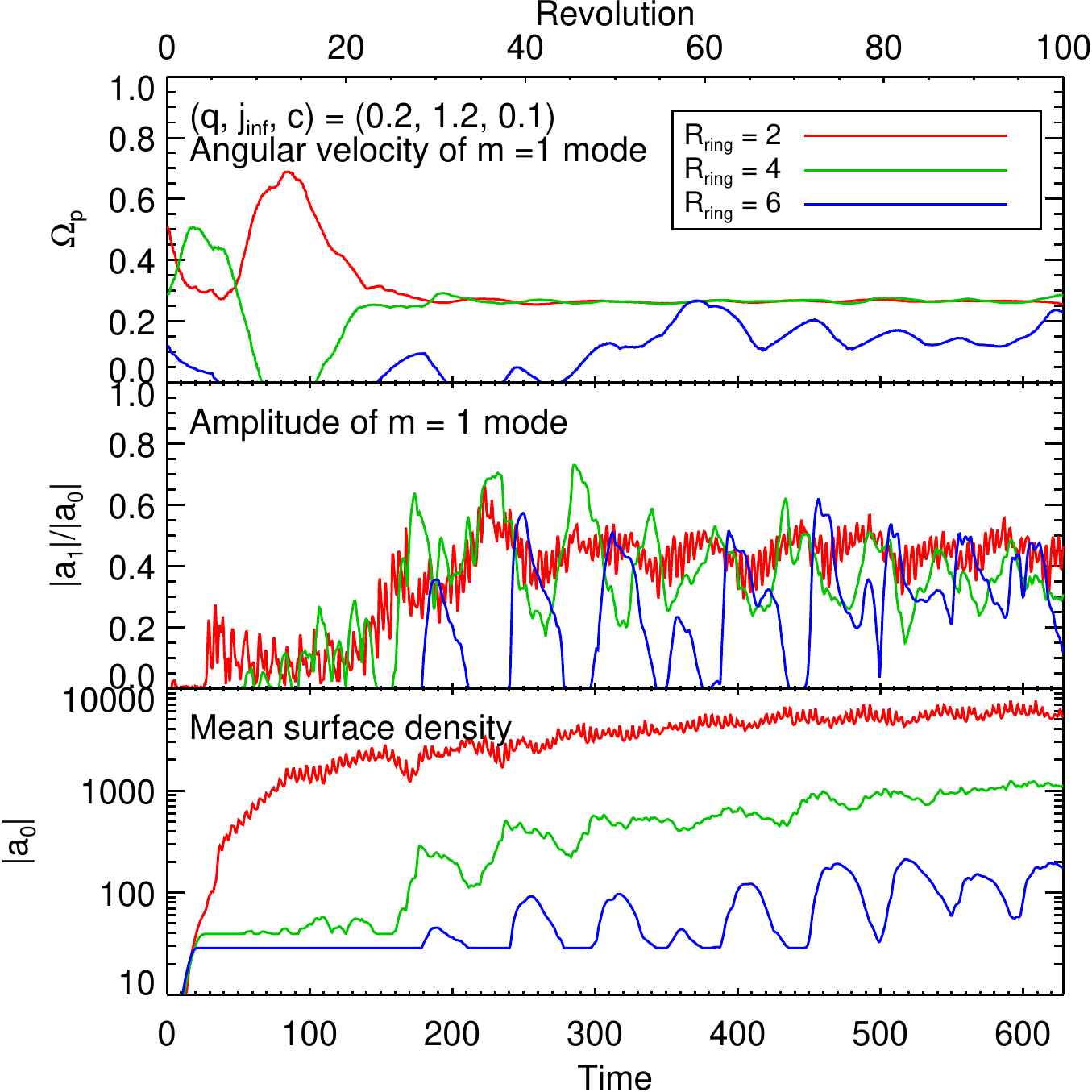}
\caption{
 Angular velocity of the $m=1$ mode $\Omega_p$ (top), amplitude of the $m=1$ mode $|a_1|/|a_0|$ (middle), and mean surface density $|a_0|$ (bottom) as a function of time. The red, green, and blue lines are for the values that were measured in the cylindrical shells with $R_\mathrm{ring} = 2$, 4, and 6. The upper abscissa shows the time in units of the rotation period of the binary.
 \label{disk_oscillate_diagram2.pdf}
 }
\end{figure}

In order to evaluate the asymmetry of the circumbinary disk, we measured the surface density in thin cylindrical shells with radii of $R = R_\mathrm{ring}\pm 0.1 $ for $R_\mathrm{ring} = 2, 4, 6$ for each stage, as shown in Figure~\ref{disk_oscillate_diagram1.pdf}. The innermost shell ($R_\mathrm{ring} = 2$) has the largest asymmetry amplitude among the three shells, thus indicating that the asymmetry is larger in the inner region of the circumbinary disk. The inclined stripes in the figure indicate that the asymmetry rotates at an almost constant angular velocity. Note that the slope of the stripes coincides with the angular velocity of the asymmetry.

In order to more quantitatively analyze the asymmetry of the circumbinary disk, the Fourier transform is performed on the surface density in the azimuthal distribution for each stage, yielding the Fourier components of the $m$-th mode, $a_m$. The amplitudes of the $m$-th modes, $|a_m|$, and their phases (or position angles), $\arg(a_m)$, are therefore evaluated. The time differential of the phase coincides with the angular velocity of the $m$-th mode.

The amplitude of $m=0$ mode, $|a_0|$, denotes the azimuthal mean surface density (Figure~\ref{disk_oscillate_diagram2.pdf}, bottom panel). The mean surface density at $R_\mathrm{ring} = 2$ increases almost linearly. Note that the ordinates are on a logarithmic scale. The radial growth of the circumbinary disk is also seen in the mean values of the surface density at $R_\mathrm{ring} = 4$ and 6. These values also shows the oscillation with a period of $\sim 10$~revolutions, indicating that the circumbinary disk oscillates radially with that period. The oscillation at $R_\mathrm{ring} = 6$ is reflected by the oscillation of the outer edge of the circumbinary disk.

The ratio of $|a_1|/|a_0|$ denotes the relative amplitude of the asymmetry (Figure~\ref{disk_oscillate_diagram2.pdf}, middle panel). In $t \gtrsim 200$, the relative amplitude reaches $\sim 50\%$ for the surface density at $R_\mathrm{ring} = 2$ and 4 with considerable oscillation, thereby indicating that the azimuthal contrast is $\sim 3\, (=150\%/50\%)$ at most. The $m=1$ mode has the largest amplitude among the modes with $m \ge 1$.

The angular velocity of the asymmetry is measured by $\Omega_p = \left[\phi_1(t+5T_\star)-\phi_1(t-5T_\star) \right]/(10 T_\star)$, which gives a temporal mean of the angular velocity over 10 revolutions ($10 T_\star$), where $\phi_1(t) = \arg\left[a_1(t)\right]$ (Figure~\ref{disk_oscillate_diagram2.pdf}, top panel). Note that the angular velocity $\Omega_p$ is for the asymmetric pattern and not for the gas velocity. The angular velocity of the asymmetry exhibits a constant value when the relative amplitude has a significant value; $\Omega_p \sim 0.25\Omega_\star$ in $t \gtrsim 200$ for $R_\mathrm{ring}=2$ and 4. This angular velocity suggests that the rotation of the asymmetric pattern is caused by resonance with the binary orbit, showing $\Omega_p: \Omega_\star \simeq 1:4$. More precisely, the mean angular velocity of the asymmetry is measured as $\Omega_p = 0.266 \Omega_\star$ in the period of $80-100$ revolutions for $R_\mathrm{ring} = 2$, as shown in the fourth column of Table~\ref{table:model-parameters}. Thus, the asymmetry has a corotation radius of $R = 2.42$ when assuming the Keplerian rotation. The dependence of $\Omega_p$ on the model parameters is shown in Section \ref{sec:dependence_c} and \ref{sec:dependence_j}.

\subsection{Vortex in the Circumbinary Disk}
\label{sec:vortex}
\begin{figure*}
\epsscale{1.1}
\plotone{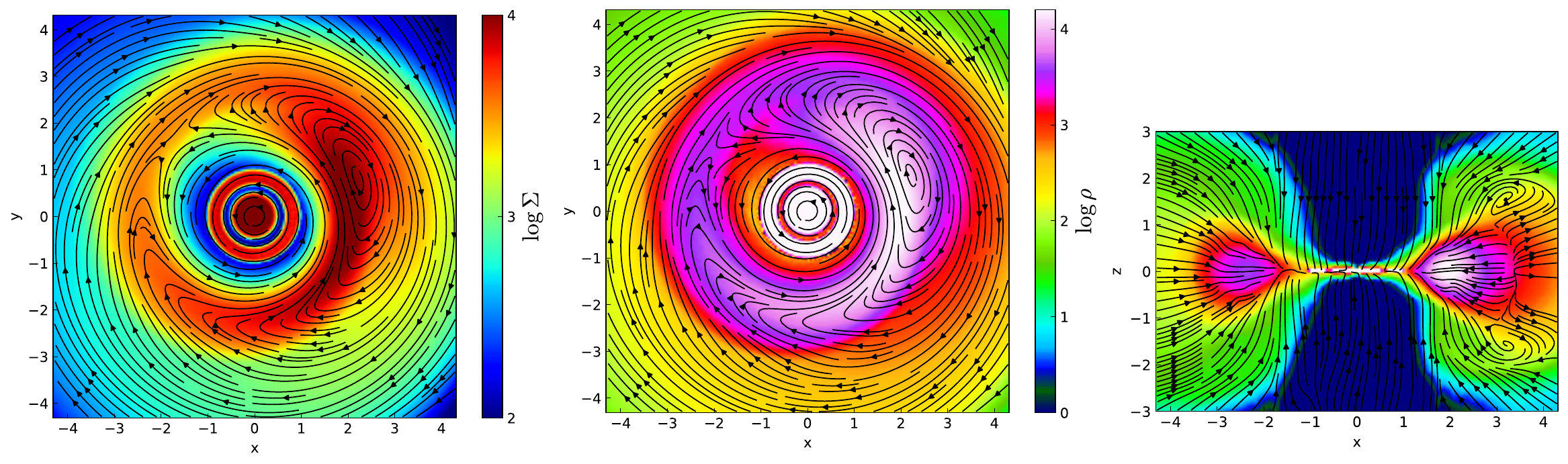}
\caption{
 Temporal average of the circumbinary disk in the period of $80 - 100$ revolutions of the binary for the model with $(q, j_\mathrm{inf}, c) = (0.2, 1.2, 0.1)$. The temporal average is calculated in the rotating frame with the angular velocity of $0.266 \Omega_\star$, which is the mean pattern speed of the asymmetry mode of the circumbinary disk in the period of 80 - 100 revolutions. The color scales depict the column density distribution (left panel), the density distribution in the $z=0$ plane (middle panel), and the $y=0$ plane (right panel). The streamlines indicate the density-weighted velocity distribution (left panel), the velocity distribution in the $z=0$ plane (middle panel), and the $y=0$ plane (right panel). In all the panels, the frame is rotated so that the maximal surface density lies on the $x$-axis. The angular velocity of the frame is different from those of the binary, and two circumstellar disks are shown as a central disk and a ring in the edge-on views.
\label{rbinAcc_p02j12c01v10_ub7215464.pdf}
}
\end{figure*}

The rotating asymmetric pattern causes the vortex in the circumbinary disk. Figure~\ref{rbinAcc_p02j12c01v10_ub7215464.pdf} shows temporal averages of the velocity, surface density, and density distributions. The temporal averages are calculated in the rotating frame with the angular velocity $\Omega_p$, which is the angular velocity of the asymmetric pattern. As shown in the face-on view (Figure~\ref{rbinAcc_p02j12c01v10_ub7215464.pdf}, left, and middle panels), the asymmetry is clearly seen in both the surface density and density distributions. The peak surface density and density are, respectively, $\sim 4$ and $\sim 6$ times larger than those on the opposite sides, at the stages shown here. In the figure, the position of the density peak nearly coincides with that of the surface density, $(x, y) \simeq (-2, 0)$.

The streamlines show the vortex, the center of which nearly coincides with the surface density and density peaks, but it is slightly shifted in the rotational direction. The aspect ratio of the vortex (the aspect ratio of the closed streamlines) is $\sim 2$. Although it has been suggested that such a thick vortex (a small aspect ratio) would be unstable \citep[e.g.,][]{Lesur09,Richard13}, the vortex here remains for a long time. This suggests that the vortex continues to be excited.

The thickness of the circumbinary disk is roughly the same between the dense and less dense sides (Figure~\ref{rbinAcc_p02j12c01v10_ub7215464.pdf}, right panel). This indicates that asymmetry in the surface density can be attributed to the density rather than the thickness of the disk.

\begin{figure*}
\epsscale{1.1}
\plottwo{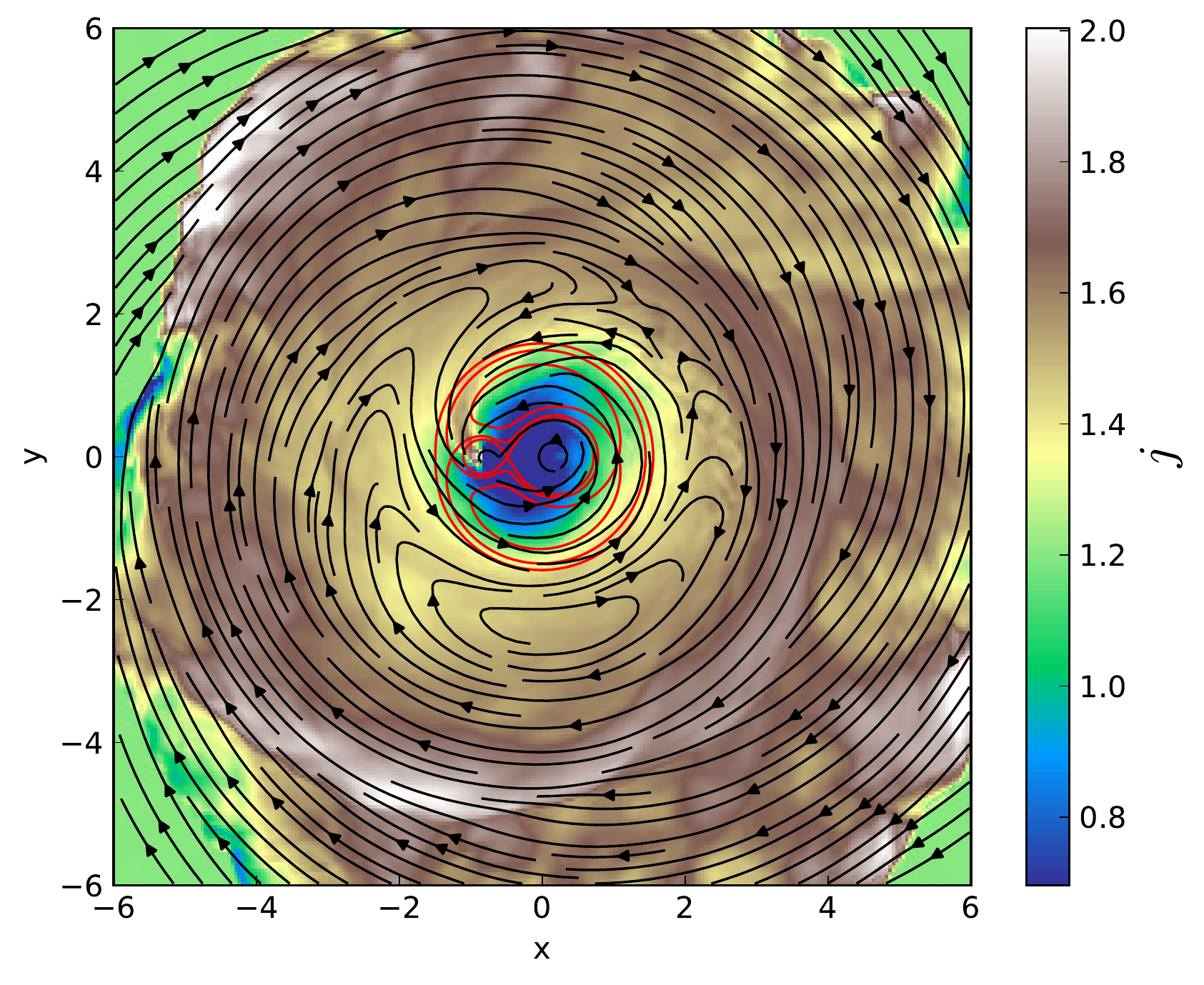}{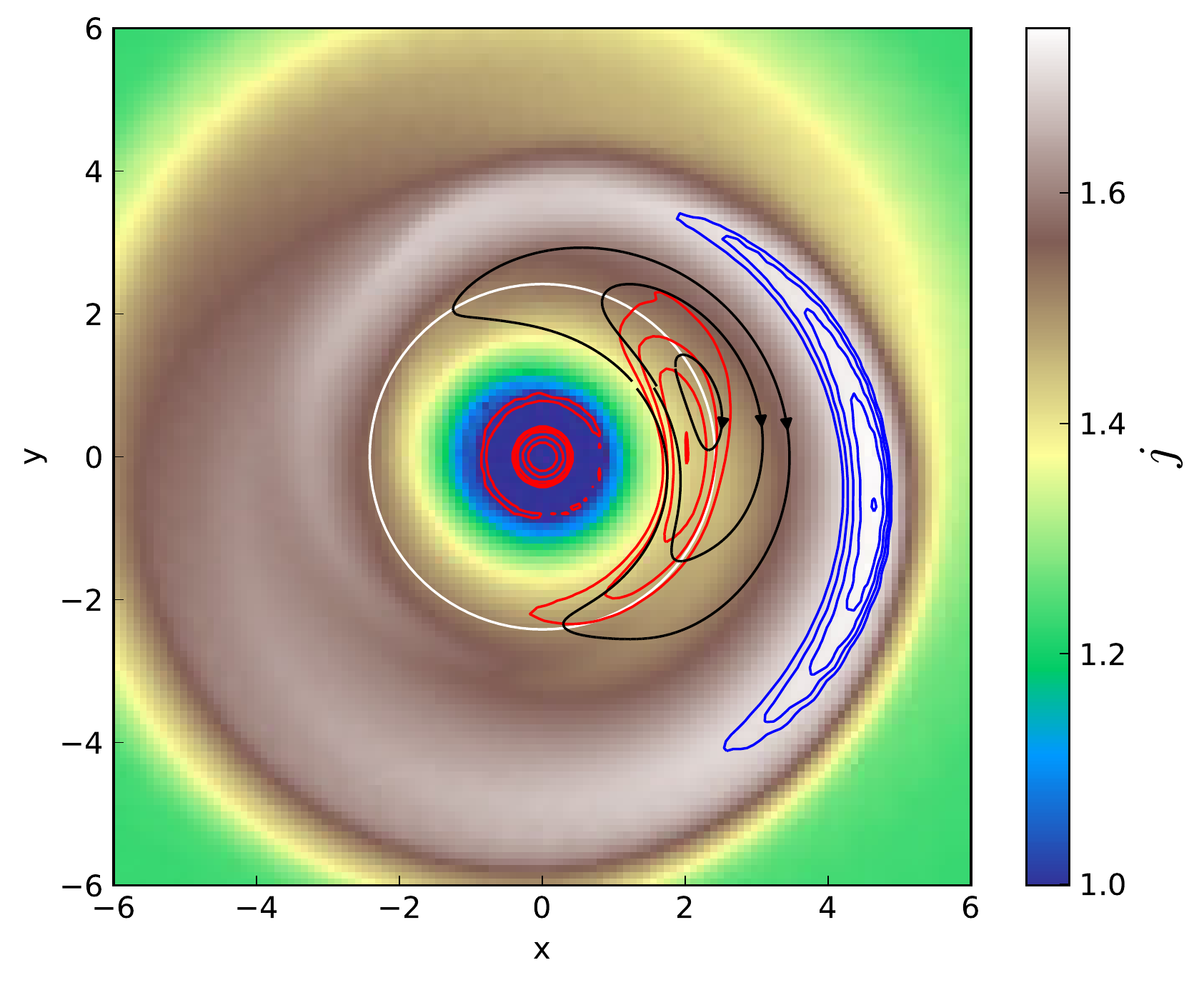}
\caption{
Specific angular momentum distribution (color) and velocity (streamlines) for the fiducial model with $(q, j_\mathrm{inf}, c) = (0.2, 1.2, 0.1)$. The left panel shows the snapshot at 100 revolutions, while the right panel shows the temporal average in the period of $80 - 100$ revolutions in the rotating frame with $\Omega_p = 0.266 \Omega_\star$. The streamlines are plotted in accordance with the velocity measured in the rotating frame with $\Omega_p$. The specific angular momentum is measured in the rest frame.  The specific angular momentum and velocity field distributions were calculated by the density-weighted average in the $z$-direction. In the left panel, the red contour lines show the Roche potential. In the right panel, the blue and red contour lines show the specific angular momentum distribution and the surface density distribution near their peaks, respectively. Their contour levels are $j= 1.70, 1.71,\cdots 1.74$, and $\Sigma = 3.8, 3.9,\cdots 4.1$. The white circle denotes the centrifugal radius of the Keplerian rotation with an angular velocity of $\Omega_p$ (corotation radius). The stage and region plotted in the left panel are same as those in Figure~\ref{rbinAcc_p02j12c01v10_arrow.pdf} (upper right). 
 \label{rbinAcc_p02j12c01v10_ug7215464_sigma.2.1.pdf}
}
\end{figure*}

The left panel of Figure~\ref{rbinAcc_p02j12c01v10_ug7215464_sigma.2.1.pdf} shows the specific angular momentum distribution on a color scale. The stage and spatial scale for this snapshot are the same as those in the upper right panel of Figure~\ref{rbinAcc_p02j12c01v10_arrow.pdf}. These figures show that the spiral arms have a high specific angular momentum. The outermost spiral arms have an especially high value of $j \gtrsim 1.8$. This high specific angular momentum is attributed to the gravitational torque of the binary. Just inside this spiral arm, a vortex is excited, as shown in the streamlines around $(x, y) \sim (0, -2)$. The surface density bump is associated with the vortex at roughly the same position (see Figure~\ref{rbinAcc_p02j12c01v10_arrow.pdf}).

This structure is observed clearly when we take a temporal average of the disk (Figure~\ref{rbinAcc_p02j12c01v10_ug7215464_sigma.2.1.pdf}, right panel). The center of the vortex (black streamlines) nearly coincides with the density bump (red contours). It is located at the corotation radius with the angular velocity of the asymmetry $\Omega_p$ for the Keplerian rotation (see white circle). We also found that the peak of the specific angular momentum $j$ (blue contours) is located in almost the same orientation  as the density peak and the center of the vortex with respect to the origin and confirmed that this configuration is maintained in the other periods, i.e., the $30-50$ and $50-70$ revolutions periods. This indicates that the density bump and vortex are caused by the locally enhanced angular momentum, which results from the gravitational torque. Further discussion of the asymmetry is presented in section~\ref{sec:discussion_asymmetry}.

\subsection{Dependence on Temperature}
\label{sec:dependence_c}

\begin{figure*}
\epsscale{1.1}
\plotone{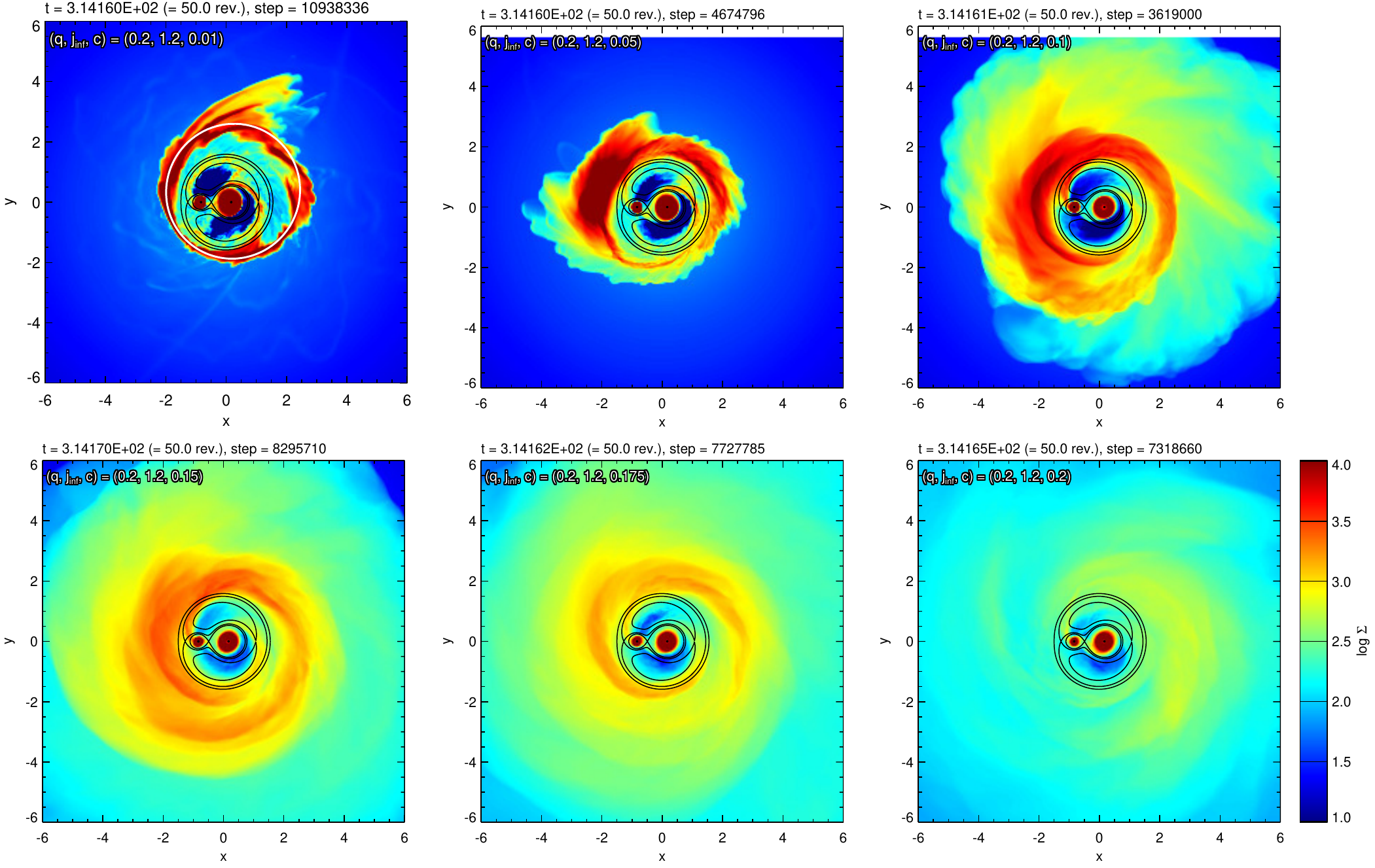}
\caption{
Surface density distributions for the models with $(q, j_\mathrm{inf}) = (0.2, 1.2)$, and $c=0.01, 0.05, 0.1, 0.15, 0.175, 0.2$ from upper left to lower right, at 50 revolutions of the binary. In each panel, the Roche potential is shown by the contour, the levels of which are the potential levels of the L1, L2, and L3 Lagrange points. In the upper left panel, the white ellipse denotes the fitted elliptic orbit. \label{rbinAcc_p02j12cxxv10_ug.pdf}
}
\end{figure*}

Figure~\ref{rbinAcc_p02j12cxxv10_ug.pdf} shows the dependence of the surface density distribution on the temperature of the gas or sound speed. The cold models (small $c$) have circumbinary disks with narrow ring shapes, while the hot models (large $c$) have circumbinary disks with large radii. Moreover, a colder disk exhibits sharper contrast in the surface density distribution. These temperature dependencies are due to changes in the thermal pressure.
 
\begin{figure}
\epsscale{1.1}
\plotone{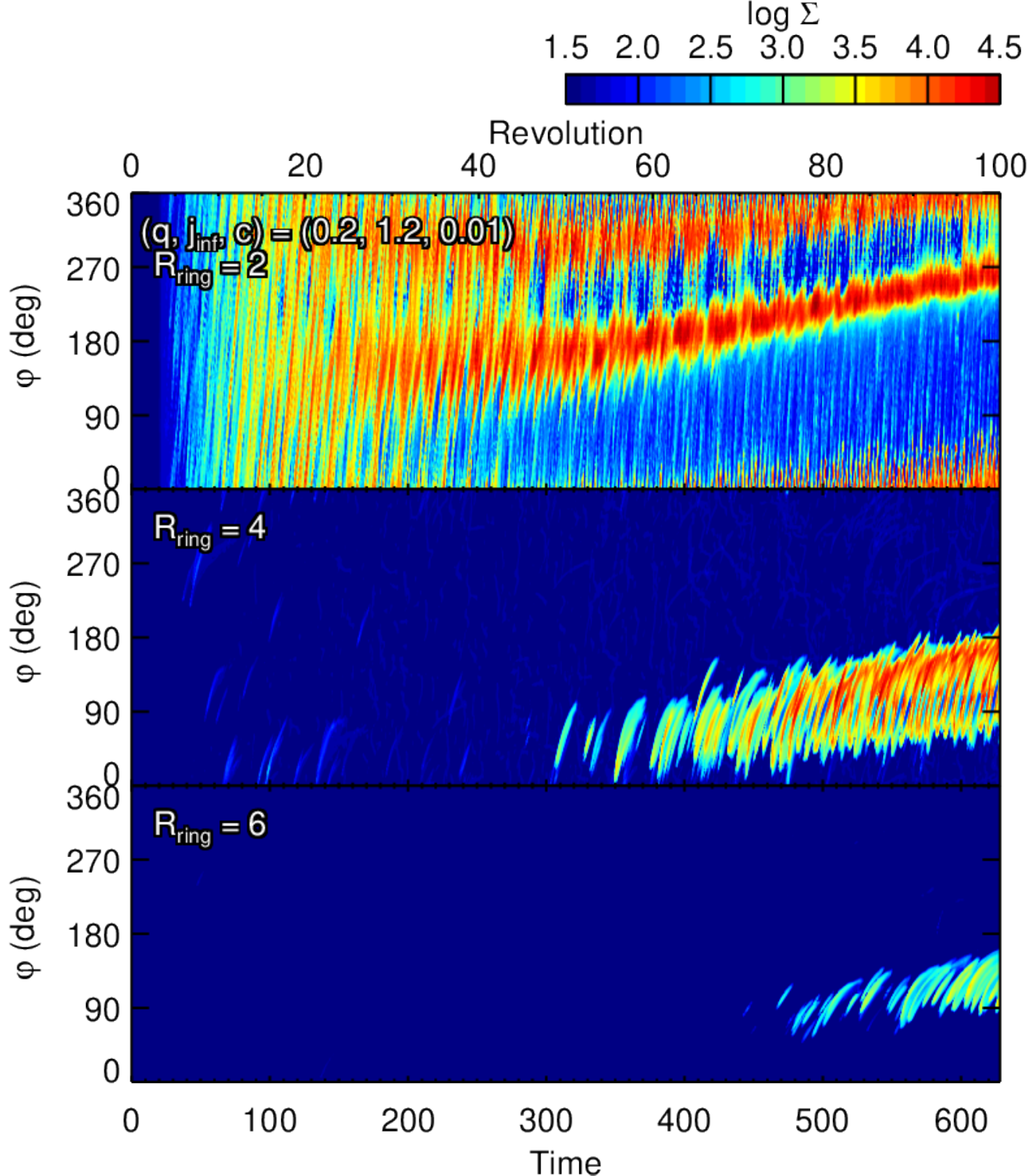}
\caption{
 Same as Figure~\ref{disk_oscillate_diagram1.pdf} but for the cold model with $(q, j_\mathrm{inf}, c) = (0.2, 1.2, 0.01)$
 \label{disk_oscillate_diagram_rbinAcc_p02j12c001v10_1.pdf}
 }
\end{figure}

\begin{figure}
\epsscale{1.1}
\plotone{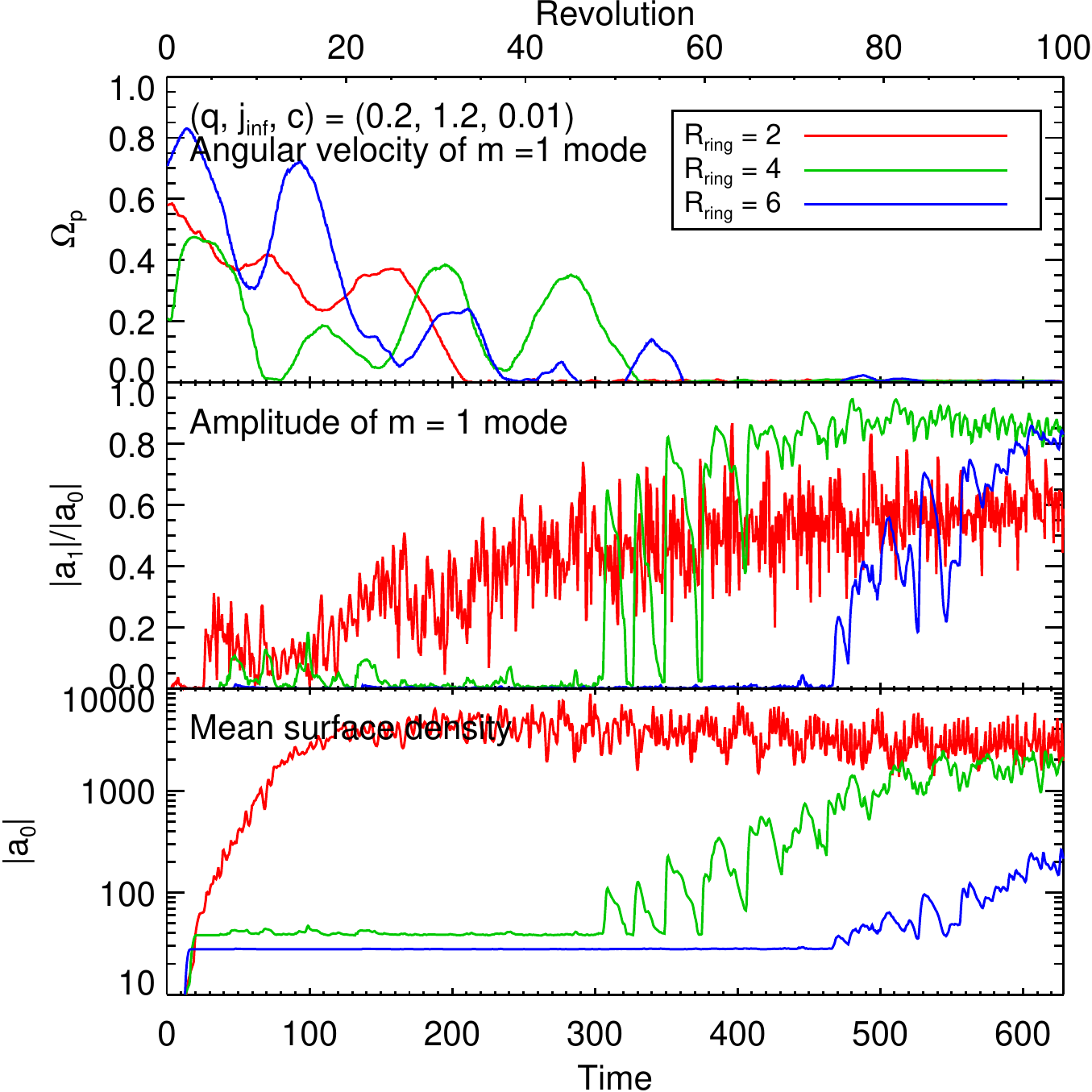}
\caption{
 Same as Figure~\ref{disk_oscillate_diagram2.pdf} but for the cold model with $(q, j_\mathrm{inf}, c) = (0.2, 1.2, 0.01)$
 \label{disk_oscillate_diagram_rbinAcc_p02j12c001v10_2.pdf}
 }
\end{figure}


The cold models also have asymmetric surface density distributions in the circumbinary disks. The model with $(q, j_\mathrm{inf}, c) = (0.2, 1.2, 0.01)$ has a narrow circumbinary disk that is off-center with respect to the barycenter of the binary stars (upper left panel of Figure~\ref{rbinAcc_p02j12cxxv10_ug.pdf}). While this circumbinary disk is almost concentric in the early phase, it becomes eccentric at $\sim 50$~revolutions of the binary. The narrow circumbinary disk is fitted by an elliptic orbit, the focus of which coincides with the barycenter of the binary stars (white ellipse in the upper left panel of Figure~\ref{rbinAcc_p02j12cxxv10_ug.pdf}). The elliptic orbit is parameterized by the angular momentum, eccentricity, and periapsis. The angular momentum was obtained by the density-weighted average in the circumbinary disk, and the eccentricity and periapsis were obtained by the least mean squares method. The angular momentum and the eccentricity gradually increase as time proceeds because of the gravitational torque of the stars. The periapsis of the circumbinary disk rotates slowly. The angular velocity is $\Omega_p = 0.00504 \Omega_\star$ in the period of $80-100$ revolutions, as shown in Table~\ref{table:model-parameters}. This low angular velocity is also shown in Figure~\ref{disk_oscillate_diagram_rbinAcc_p02j12c001v10_1.pdf}, in which the peak surface density diagrams shows a shallow slope. Note that the two peaks of the surface density in the panel of $R_\mathrm{ring} = 2$ correspond to the two intersections between the eccentric circumbinary disk and the cylindrical ring with $R_\mathrm{ring} = 2$. Figure~\ref{disk_oscillate_diagram_rbinAcc_p02j12c001v10_2.pdf} shows the low $\Omega_p$ in the stages where the asymmetry $|a_1|/|a_0|$ has a large amplitude. 

The slow rotation of the asymmetry in the cold models is qualitatively different from that in the fiducial model. \citet{Moriwaki04} derived a formula for the precession rate of a test particle orbiting a binary by using a secular perturbation theory, and \citet{Thun17} modified it as follow \citep[see also][]{Leung13,Dunhill15}
\begin{equation}
\Omega_p = \frac{3}{4} \frac{q}{(1+q)^2} \left(\frac{D}{a_p}\right)^{7/2} \left(1-e_p^2\right)^{-2} \Omega_\star \;,
\label{eq:Thun17}
\end{equation}
where $a_p$ and $e_p$ are the semi-major axis and eccentricity of the test particle, respectively. Equation~(\ref{eq:Thun17}) is an expression for the case of a circular orbiting binary. For the model with $c=0.01$, the surface density of the circumbinary disk is fitted to an elliptic orbit by using the method described above, yielding $a_p = 3.0$ and $e_p=0.42$ as temporal averages in the period of $80-100$ revolutions. Using these orbital parameters, Equation~(\ref{eq:Thun17}) gives $\Omega_p = 0.0032\Omega_\star$. For the model with $c=0.02$, we obtained $a_p=2.9$, $e_p = 0.40$, and $\Omega_p = 0.0033\Omega_\star$. These values are consistent with the value listed in Table~\ref{table:model-parameters}, indicating that the secular motion of the ballistic particle causes the rotation of the asymmetry of the circumbinary disk and the hydrodynamical effects are minor in the cold models. 

\begin{figure}
\epsscale{1.1}
\plotone{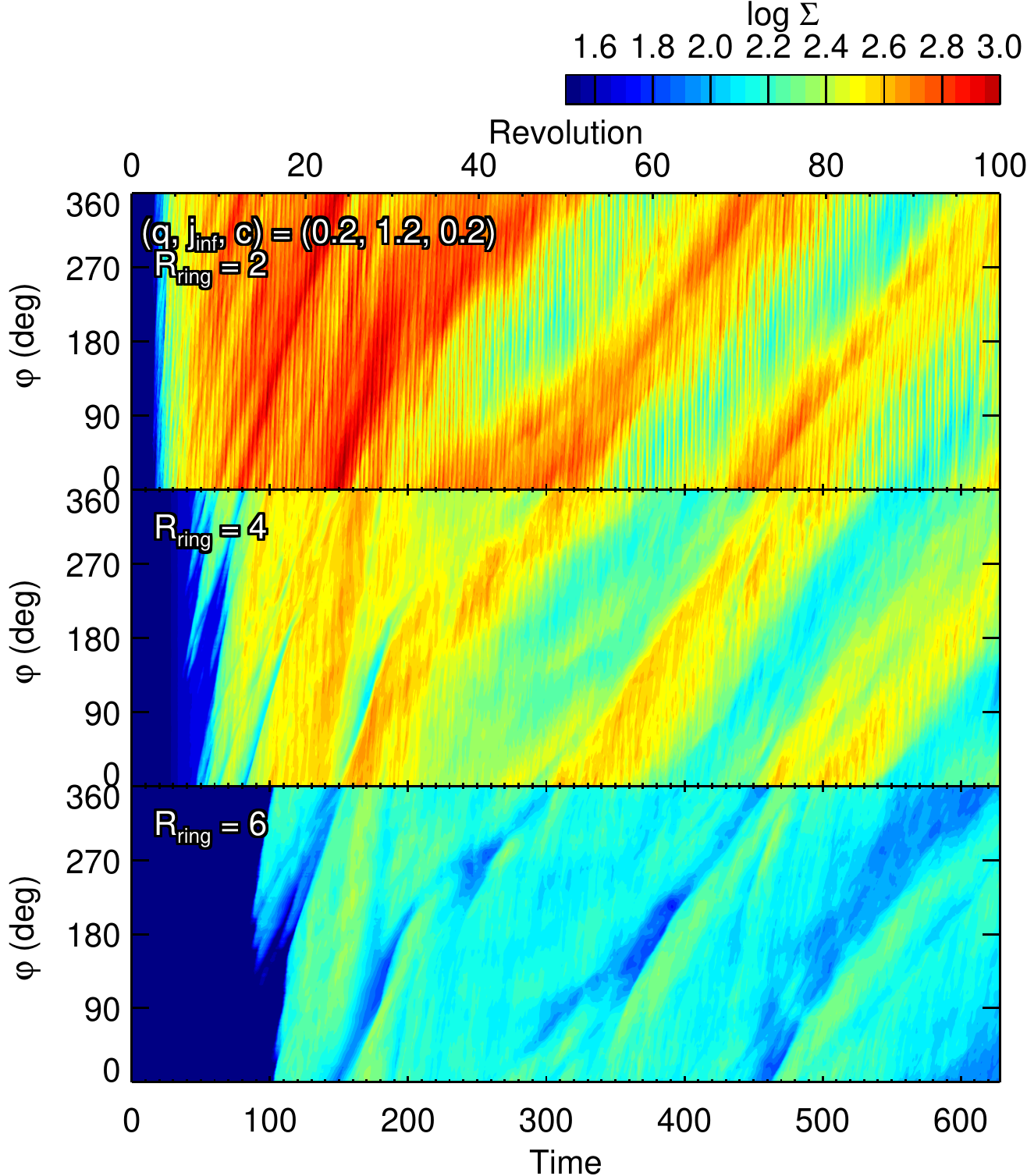}
\caption{
 Same as Figure~\ref{disk_oscillate_diagram1.pdf} but for the hot model with $(q, j_\mathrm{inf}, c) = (0.2, 1.2, 0.2)$
 \label{disk_oscillate_diagram_rbinAcc_p02j12c02v10_1.pdf}
 }
\end{figure}

\begin{figure}
\epsscale{1.1}
\plotone{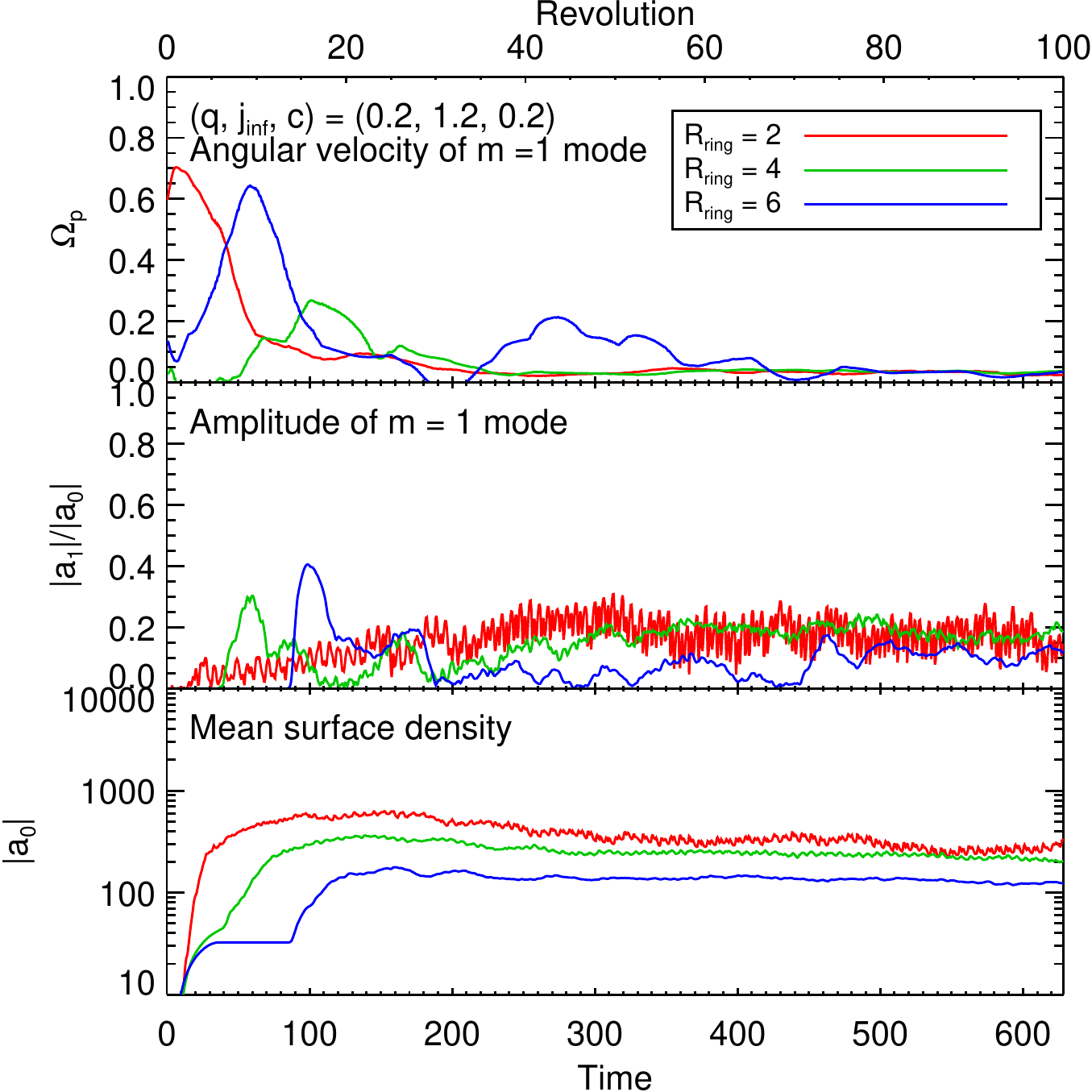}
\caption{
 Same as Figure~\ref{disk_oscillate_diagram2.pdf} but for the hot model with $(q, j_\mathrm{inf}, c) = (0.2, 1.2, 0.2)$
 \label{disk_oscillate_diagram_rbinAcc_p02j12c02v10_2.pdf}
 }
\end{figure}


The hot models show the circumbinary disks extended up to large radii as shown in the lower panels of Figure~\ref{rbinAcc_p02j12cxxv10_ug.pdf}. A circumbinary disk extends faster with a higher sound speed. Half of the outer edge of the circumbinary disk touches the computational boundary at $R = 12D$ at the stages of 80, 50, and 31 revolutions for the model with $c=0.15$, 0.175 and 0.2, respectively. 
For models with larger sound speed, both the surface density contrast and the amplitude of the asymmetry are lower. The extension of the circumbinary disks and the low contrast are responsible for the high pressure seen in the hot models. 

For the model with the largest sound speed ($c=0.2$), the amplitude of the asymmetry is as low as $|a_1|/|a_0|\sim 0.2$ (Figure~\ref{disk_oscillate_diagram_rbinAcc_p02j12c02v10_2.pdf}). The angular velocity of the asymmetric pattern is a very low value of $\Omega_p = 0.0335 \Omega_\star$ in the period of $80-100$ revolutions (see also Table~\ref{table:model-parameters}). The corresponding corotation radius is $R = 9.62$. This slow rotation is observed in the slope of the diagrams in Figure~\ref{disk_oscillate_diagram_rbinAcc_p02j12c02v10_1.pdf}. 

The accretion rates are affected by the temperature. All the models show a tendency to decrease in $\Gamma$ with time, while the cold models with $c \le 0.02$ show a rapid decrease in $\Gamma$, resulting in low $\Gamma$ at the ends of the calculations (Table~\ref{table:model-parameters}). The decrease in $\Gamma$ is mainly attributed to an increase in $\dot{M}_1$, as observed in the fiducial model (Figure~\ref{plot_pmass.pdf}). \citet{Young15a} reported that a model with hot gas ($c=0.25$) shows variability of the accretion rates, the timescale of which was $t \sim 90$. Our model with $c=0.2$ also exhibits variability mainly in the accretion rate of the secondary, and the timescale is longer ($t \sim 100-200$) than that of \citet{Young15a}.

\subsection{Dependence on Angular Momentum of Infalling Gas}
\label{sec:dependence_j}

\begin{figure*}
\epsscale{1.1}
\plotone{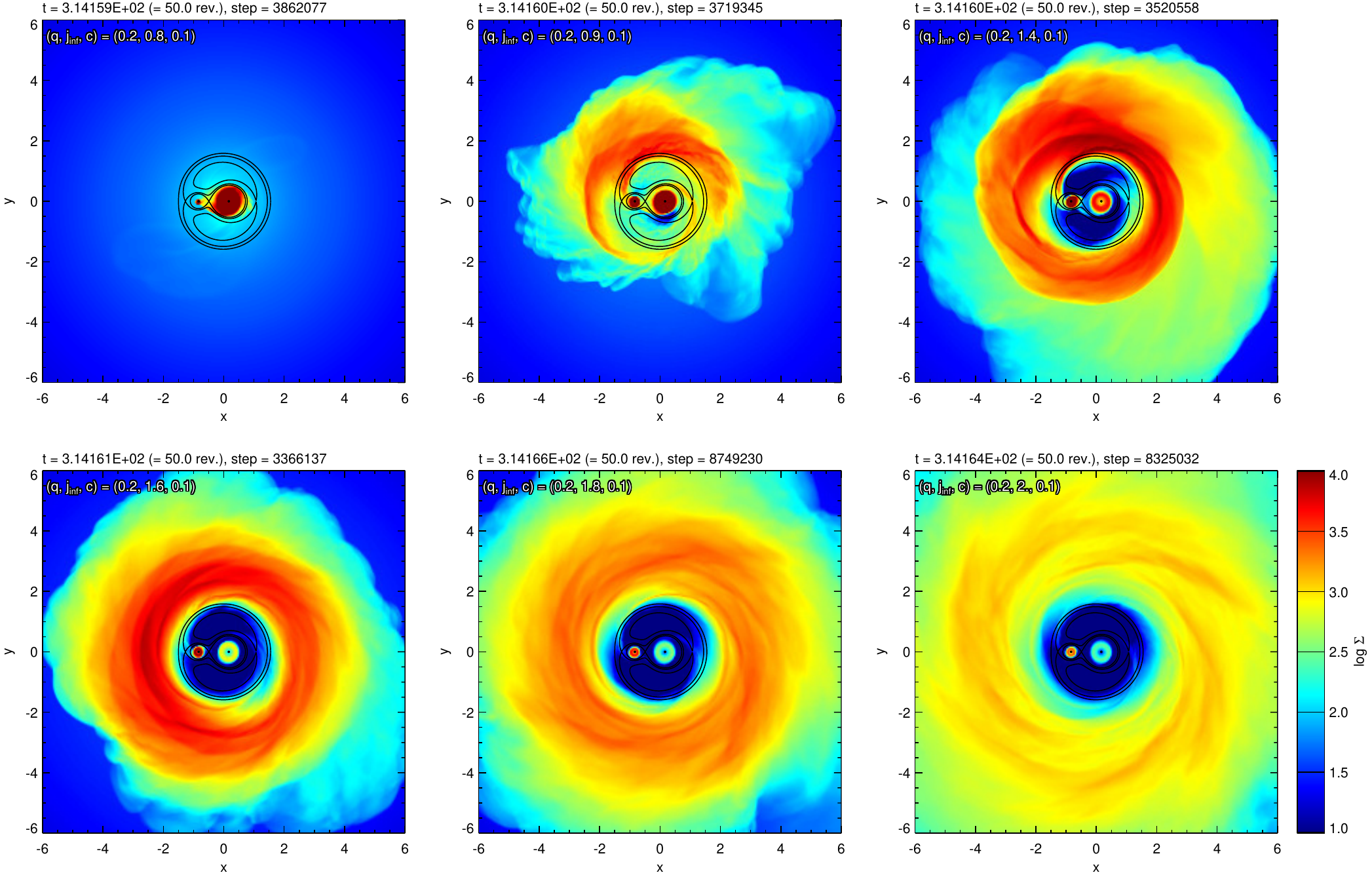}
\caption{
Surface density distributions for the models with $(q, c) = (0.2, 0.1)$, and $j_\mathrm{inf}=0.8, 0.9, 1.4, 1.6, 1.8, 2.0$ from the upper left to lower right, at 50 revolutions of the binary. In each panel, the Roche potential is shown by contours, the levels of which are the potential levels of the L1, L2, and L3 Lagrange points.
 \label{rbinAcc_p02jxxc01v10_ug.pdf}
}
\end{figure*}

Figure~\ref{rbinAcc_p02jxxc01v10_ug.pdf} shows the representative models with different specific angular momenta for the infalling gas. We observed that the models with $j_\mathrm{inf} \le 0.8$ do not have circumbinary disks, while models with $j_\mathrm{inf} \ge 0.9$ have circumbinary disks. This is in agreement with the results of the SPH simulations by \citet{Bate97}. They have implicitly used the condition for circumbinary disk formation, in which the centrifugal radius of the gas, $R_\mathrm{cent} = j_\mathrm{inf}^2/(GM_\mathrm{tot})$, is larger than the orbital radius of the secondary star, $a_2 = D/(1+q)$, yielding
\begin{equation}
j_\mathrm{inf} \ge \left(\frac{GM_\mathrm{tot}D}{1+q}\right)^{1/2}\;.
\label{eq:cbd_condition}
\end{equation}
For $q=0.2$, equation~(\ref{eq:cbd_condition}) gives $j_\mathrm{inf} \ge 0.91$, which is in good agreement with our simulations.

The models with the circumbinary disk ($j_\mathrm{inf} \ge 0.9$) show $\dot{M}_2 > \dot{M}_1$ throughout the evolution. On the other hand, the model with $j_\mathrm{inf} = 0.8$ shows $\dot{M}_1 > \dot{M}_2$ after $\sim 50$ revolutions. The model with $j_\mathrm{inf} = 0.5$ shows $\dot{M}_1 > \dot{M}_2$ throughout the evolution and has a negative $\Gamma$ (Table~\ref{table:model-parameters}). The dependence of $j_\mathrm{inf}$ on the accretion rates is consistent with \citet{Bate97}.

The models with large $j_\mathrm{inf}$ have extended circumbinary disks, as shown in the lower panel of Figure~\ref{rbinAcc_p02jxxc01v10_ug.pdf}. The surface density distribution for the circumbinary disk has a peak around the centrifugal radius specified by $j_\mathrm{inf}$ (e.g., $R_\mathrm{cnet} = 2.56, 3.24, 4.0$ for $j_\mathrm{inf} = 1.6, 1.8, 2.0$, respectively). For larger $j_\mathrm{inf}$, the circumbinary disk gap and the circumprimary and circumsecondary disks all show lower surface densities. The circumprimary disk is influenced more by the change in $j_\mathrm{inf}$ than the circumsecondary disk. The mass ratio for the circumstellar disks is $M_\mathrm{CSD}/M_\mathrm{CPD} = 1.26$ at 100 revolutions for the model with $j_\mathrm{inf} = 2.0$, and the circumsecondary disk is more massive than the circumprimary disk.

\begin{figure}
\epsscale{1.1}
\plotone{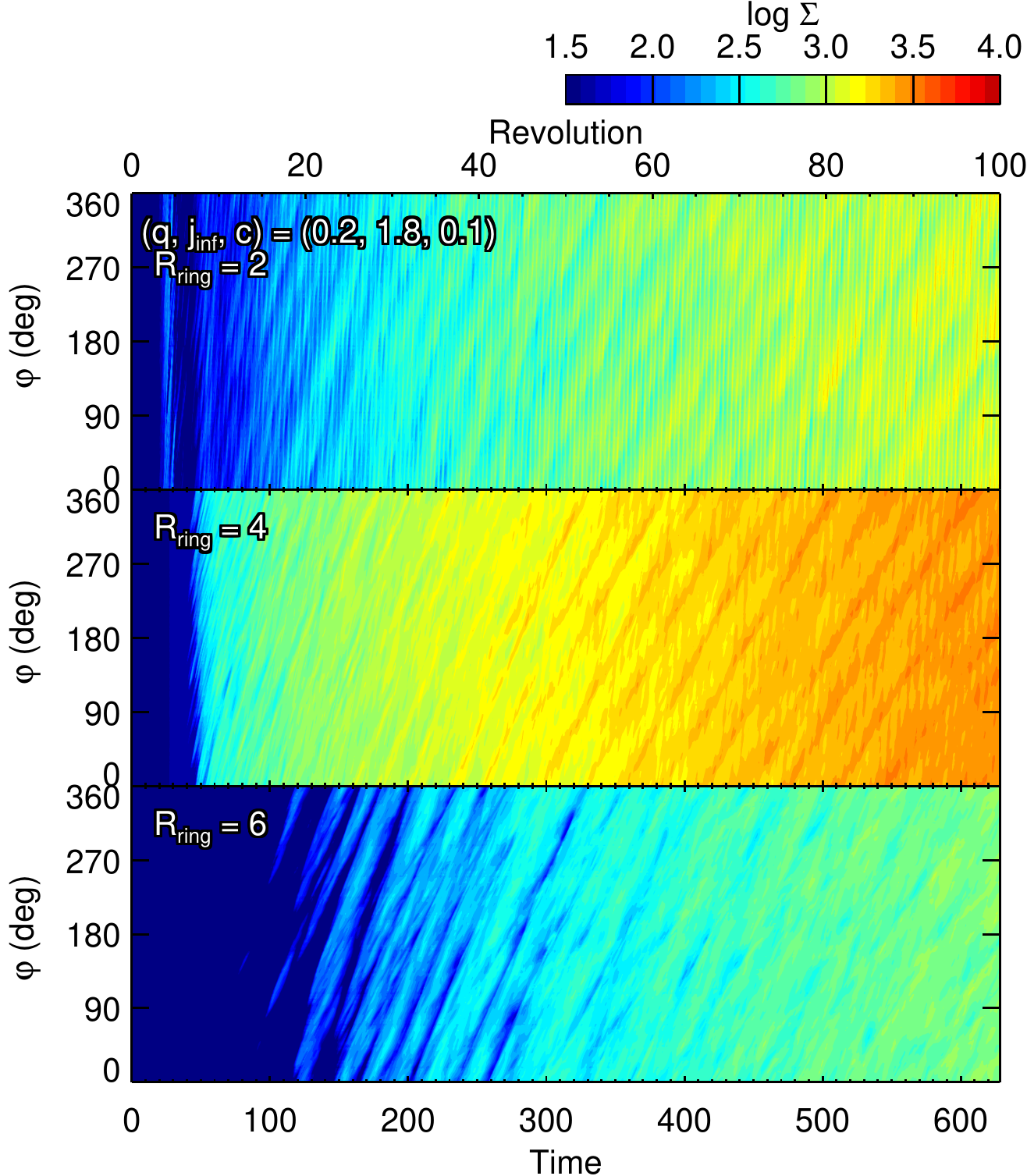}
\caption{
 Same as Figure~\ref{disk_oscillate_diagram1.pdf} but for the hot model with $(q, j_\mathrm{inf}, c) = (0.2, 1.8, 0.1)$
 \label{disk_oscillate_diagram_rbinAcc_p02j18c01v10_1.pdf}
 }
\end{figure}

The asymmetry of the circumbinary disk is qualitatively deferent when $j_\mathrm{inf} \ge 1.8$. In these models, there are several $m$-th modes with considerable amplitudes and the $m=1$ mode does not have the largest amplitude (see also Figure~\ref{disk_oscillate_diagram_rbinAcc_p02j18c01v10_1.pdf} for the model with $j_\mathrm{inf} = 1.8$). Fourier analysis shows that the amplitudes $|a_m|$ oscillate and that the largest mode with $m \ge 1$ changes with time.

\subsection{Dependence on Mass Ratio}

\begin{figure*}
\epsscale{1.1}
\plotone{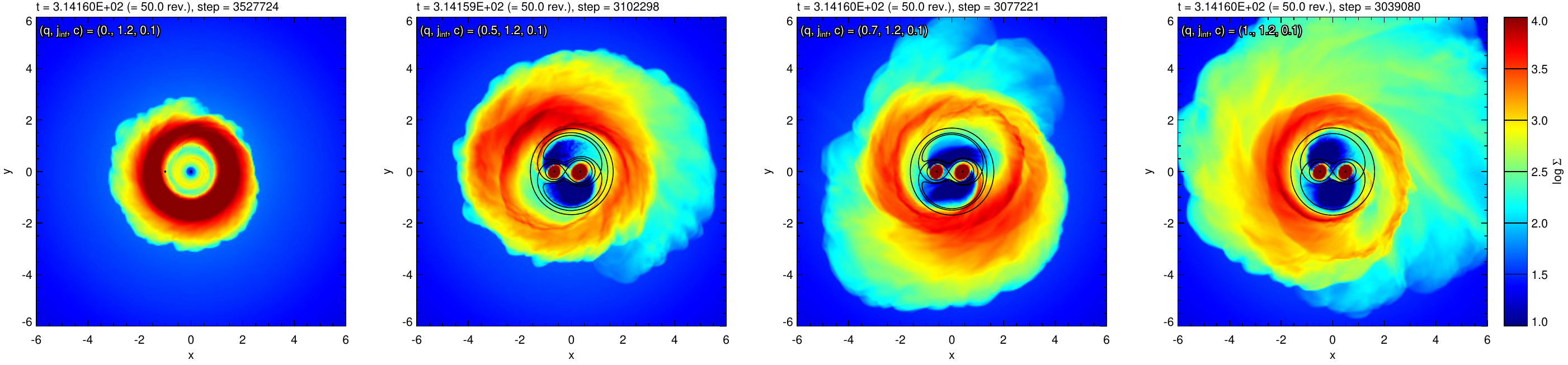}
\caption{
Surface density distribution for the models with $(j_\mathrm{inf}, c)= (1.2, 0.1)$, and $q=0.0, 0.5, 0.7, 1.0$ from left to right, at 50 revolutions of the binary. In each panel, the Roche potential is shown by contours, the levels of which are the potential levels for the L1, L2, and L3 Lagrange points. Note that the models with $q=0.0$ and $1.0$ represent the cases of a single star and an equal-mass binary, respectively. \label{rbinAcc_pxxj12c01v10_ug.pdf}
}
\end{figure*}

Figure~\ref{rbinAcc_pxxj12c01v10_ug.pdf} compares models with different mass ratios $q$. The models with $q = 0$ and 1 correspond to a single star and an equal mass binary, respectively. In the binary cases ($q \neq 0)$, the structures of the circumbinary disks are not very sensitive to the mass ratio $q$. All the binary models examined here show asymmetry in their circumbinary disks, with almost the same amplitude and angular velocity $\Omega_p$ irrespective of $q$ (see also Table~\ref{table:model-parameters}). 

For all the binary models, the radial extent of the each circumbinary disk oscillates as seen in the fiducial model. The radial extent is more sensitive to the stage than the mass ratio because of this oscillation.

The model with $q=0$ corresponds to the case of a single star (the left-most panel of Figure~\ref{rbinAcc_pxxj12c01v10_ug.pdf}). The disk has a narrow ring shape with almost constant specific angular momentum $j_\mathrm{inf}$. The central star does not cause gravitational torque, and the static gravity of the point mass acts on the circumstellar disk. We observe asymmetry of the $m=1$ mode in the ring, the amplitude of which is small compared to the binary models; $|a_1|/|a_0| \lesssim 0.2$ at $R_\mathrm{ring} = 2$. The angular velocity of the asymmetric pattern is $\Omega_p = 0.589$ (see also Table~\ref{table:model-parameters}), which coincides with that of the Keplerian rotation with the angular momentum of $j_\mathrm{inf}=1.2$. This angular velocity is considerably different from those of the binary models (the models with $q \ne 0$). Moreover, there is no significant accretion onto the star $\dot{M}_\star/\dot{M}_\mathrm{env} \lesssim 0.1\%$, where $\dot{M}_\star$ is the accretion rate and $\dot{M}_\mathrm{env}$ is the gas injection rate into the computational domain. In summary, the disk in the single star model shows different features from the circumbinary disks, indicating that the star's gravitational torque has considerable impact on the circumbinary disks.

\subsection{Effect of the Infalling Envelope}
\label{sec:effect_of_infalling_envelope}

In all the models examined above, the gas is injected from the computational boundaries, which mimics the infalling envelope. We now consider an additional model in which the gas injection at the boundaries is terminated at 50 revolutions of the binary. 

\begin{figure}
 \epsscale{1.1}
 \plotone{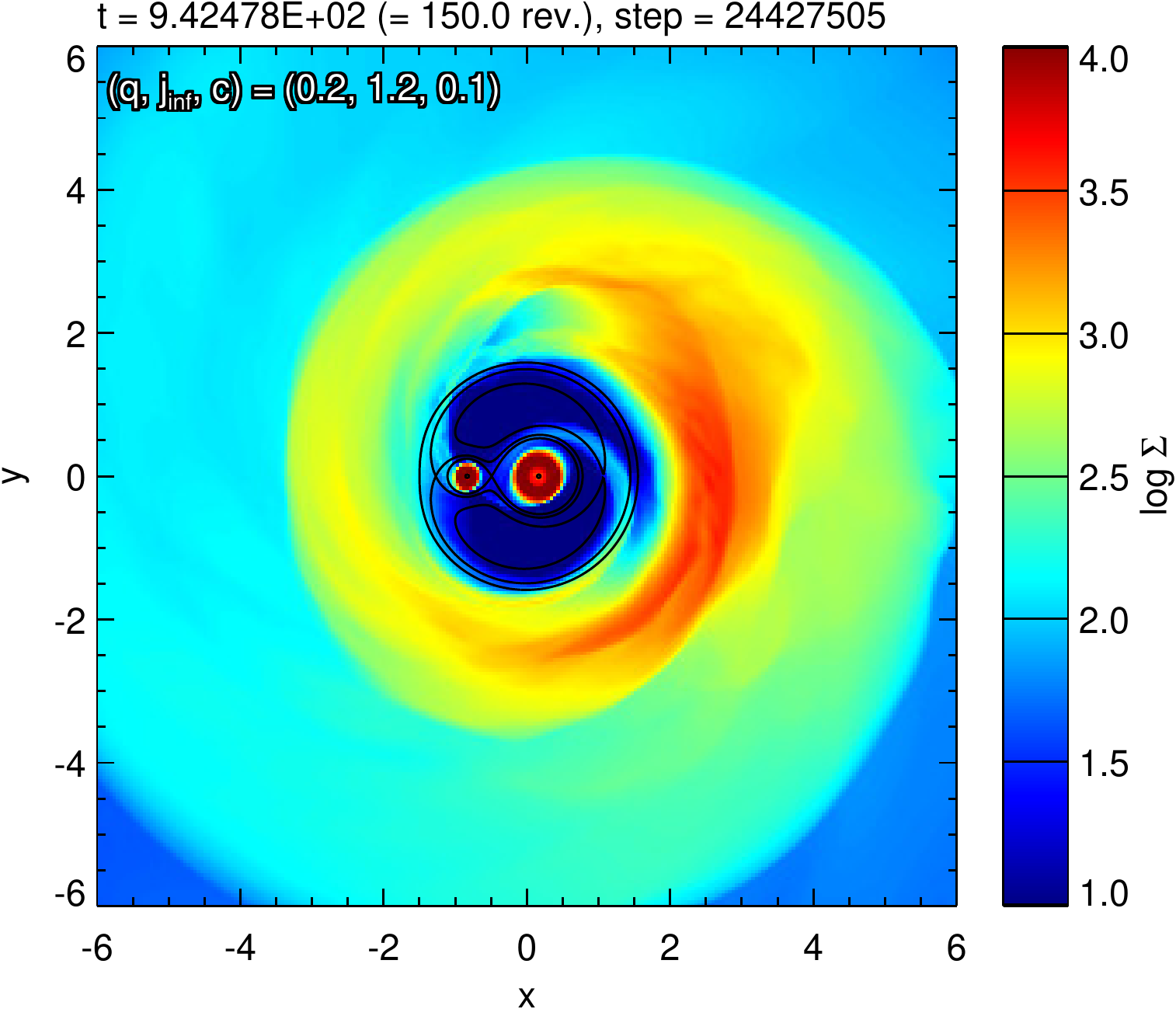}
 \caption{
Surface density distribution for the models with $(q, j_\mathrm{inf}, c)= (0.2, 1.2, 0.1)$ with gas injection only for $t = 0 - 100\pi$ (0 - 50 revolutions). A snapshot at $t=300\pi$ (150 revolutions) is shown. 
 \label{rbinAcc_p02j12c01v10-00_ug24427505.1.pdf}
 }
\end{figure}

Figure~\ref{rbinAcc_p02j12c01v10-00_ug24427505.1.pdf} shows the surface density distribution at 150 revolutions. This stage occurs 100 revolutions after the gas injection has ended. This model exhibits a smooth surface density distribution compared to the models with gas infall (compare with top right panel of Figure~\ref{rbinAcc_p02j12c01v10_arrow.pdf} and the top right panel of Figure~\ref{rbinAcc_p02j12cxxv10_ug.pdf}). This indicates that the accretion onto the circumbinary disk excites a turbulent feature there.
We also observe that the circumbinary disk extends after gas injection ends, indicating that the ram pressure of the infalling gas confines the circumbinary disk to relatively compact sizes.

The asymmetry of the circumbinary disk appears regardless of gas injection, as shown in Figure~\ref{rbinAcc_p02j12c01v10-00_ug24427505.1.pdf}. The angular velocity of the asymmetry is measured as $\Omega_p = 0.254$ and 0.235 for $80-100$ and $130-150$ revolutions, respectively. The angular velocity is insensitive to gas injection.

\begin{figure}
 \epsscale{1.1}
 \plotone{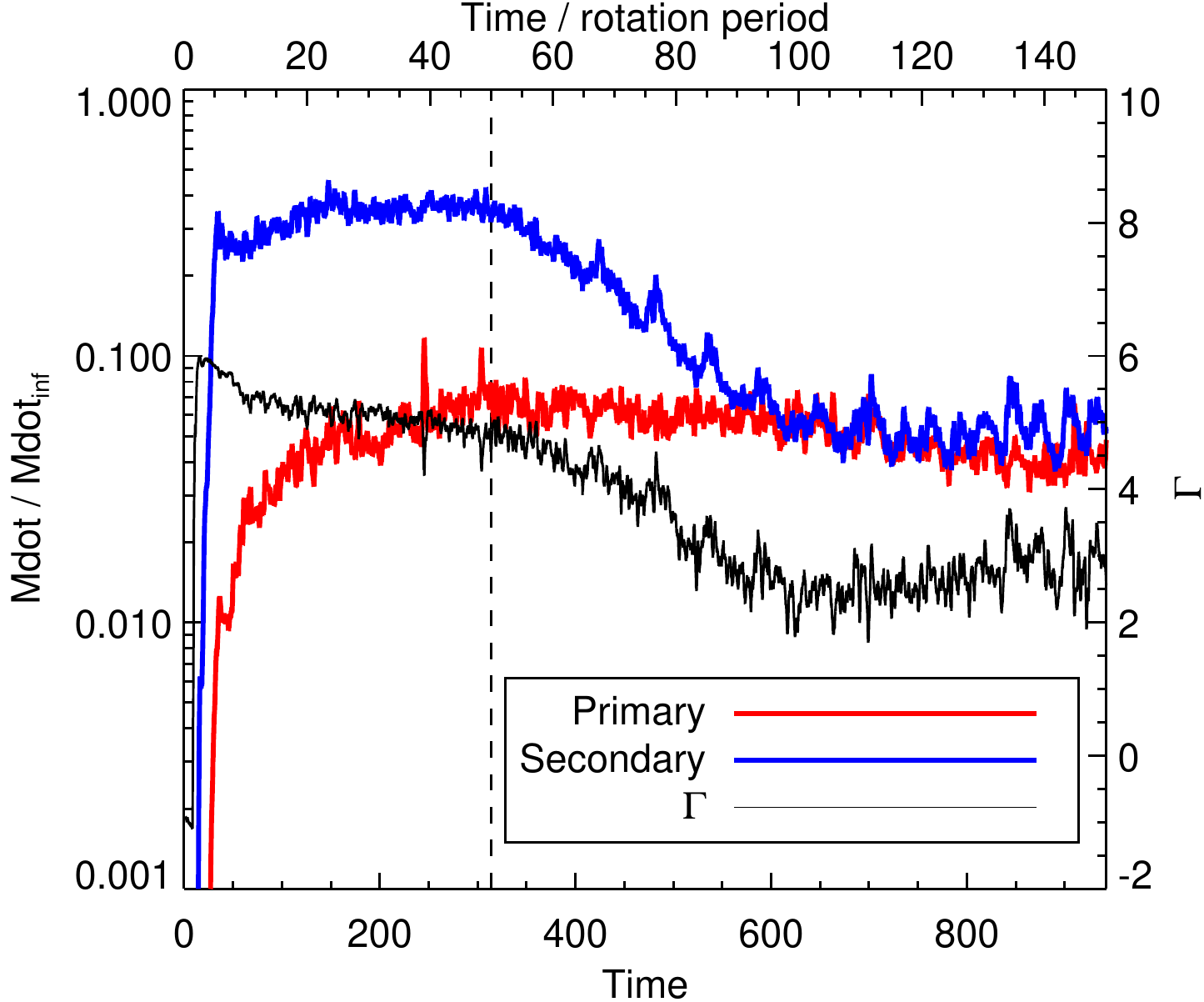}
 \caption{Accretion rates for the primary (red line) and secondary (blue line) as a function of time for the model with $(q, j_\mathrm{inf}, c) = (0.2, 1.2, 0.1)$ with gas injection for only $t = 0 - 100\pi$ (0 - 50 revolutions). The accretion rates are normalized by the mass injection rate of the boundary condition. The time in the upper abscissa is normalized by the rotation period of the binary. The change rate for the mass ratio $\Gamma$ is also shown (black line). The dashed vertical line denotes the time at which the injection ends.
 \label{plot_pmass_p02j12c01v10-00.pdf}
 }
\end{figure}

The accretion rate for each star is influenced by gas injection. As shown in Figure~\ref{plot_pmass_p02j12c01v10-00.pdf}, $\dot{M}_2$ decreases down to the level of the $\dot{M}_1$ after gas injection ends. 

\begin{figure}
 \epsscale{1.1}
 \plotone{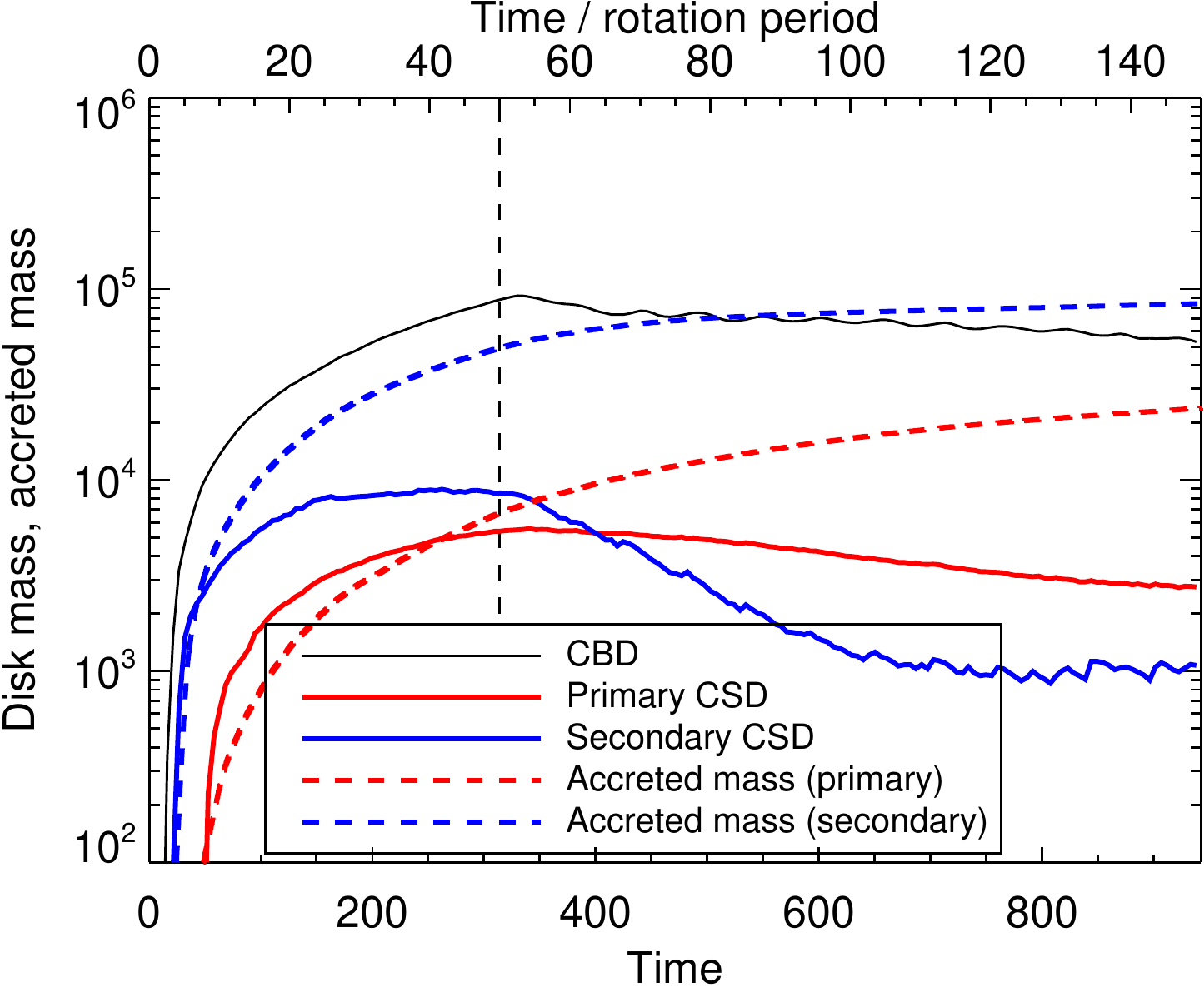}
 \caption{
 Masses of the circumbinary disk (CBD) and the circumstellar disks (CSDs) for the primary and secondary as a function of time for the fiducial model with $(q, j_\mathrm{inf}, c) = (0.2, 1.2, 0.1)$ with gas injection for only $t = 0 - 100\pi$ (0 - 50 revolutions). The dashed lines are the accreted masses to the primary and secondary for comparison. The accreted mass is defined as the accumulated mass of the gas that accretes onto each sink particle. The vertical dashed line denotes the time at which the injection ends.
 \label{disk_total_p02j12c01v10-00_log.pdf}
 }
\end{figure}

The masses of the circumbinary disk and circumstellar disks are also influenced by the termination of the gas injection. As shown in Figure~\ref{disk_total_p02j12c01v10-00_log.pdf}, the mass of the circumbinary disk decreases slowly after the gas injection ends, in contrast to the model shown in Figure~\ref{disk_total_p02j12c01v10_log.pdf}, in which the gas accretion continues. This slow decrease is due to the gas infall onto the circumstellar disks. The masses of the circumstellar disks also decrease after the gas injection ends. Especially the mass of circumsecondary disk decreases by an order of magnitude in 60 revolutions after the gas injection ends, and then it remains almost constant. 
At 150 revolutions, the ratio of the circumstellar disk is $M_\mathrm{CSD}/M_\mathrm{CPD} = 0.37$, while both the disks have almost the same mass in the model with the same parameters, but with uninterrupted gas injection. 

\begin{figure}
 \epsscale{1.1}
 \plotone{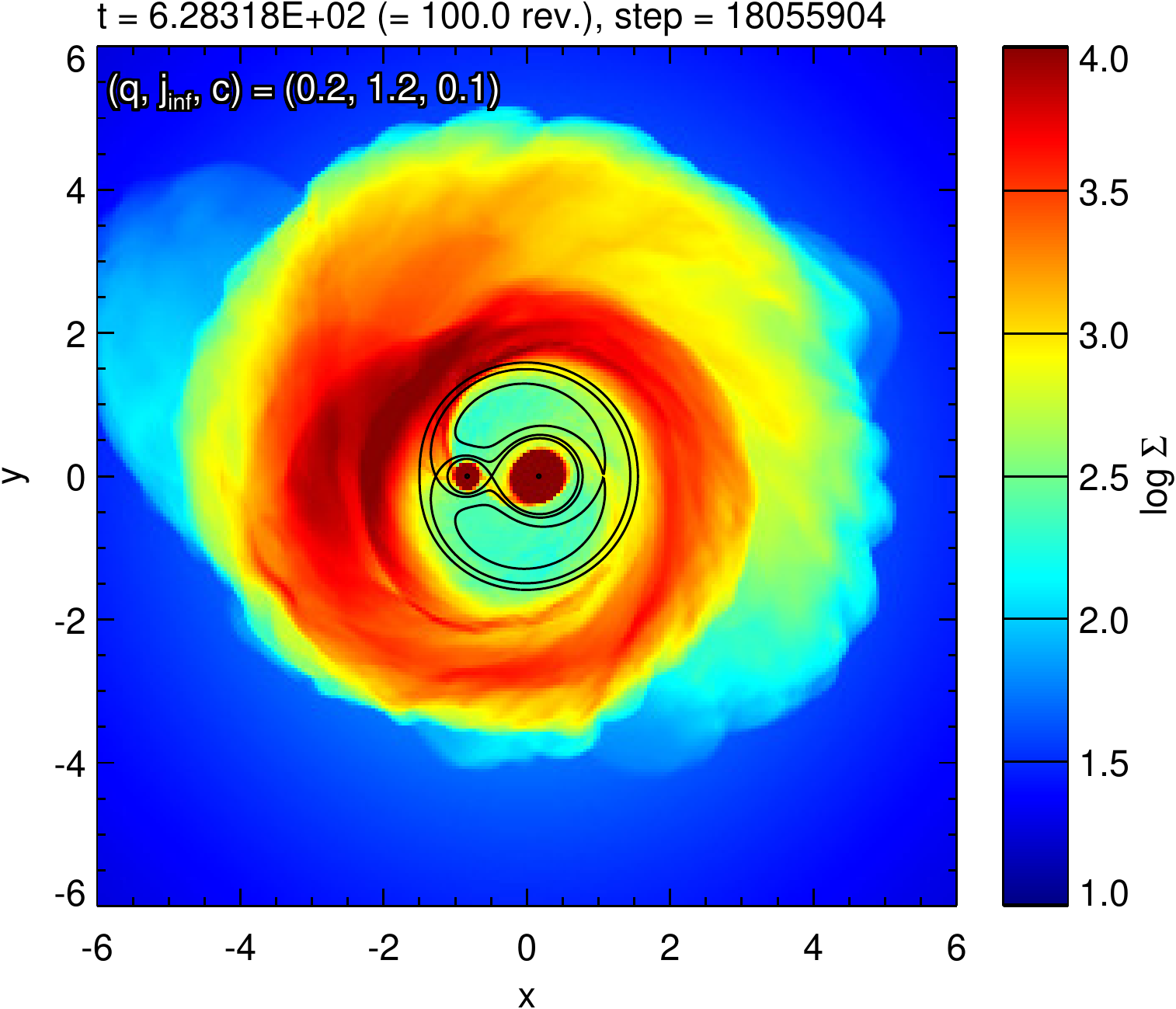}
 \caption{
 Surface density distribution for the models with the rigidly rotating envelope model of $(q, j_\mathrm{inf}, c)= (0.2, 1.2, 0.1)$ at  $t=200\pi$ (100 revolutions). 
 \label{rbinAcc_p02j12c01v10omg2_ug18055904.1.pdf}
 }
\end{figure}

In addition, we consider two rigidly rotating envelope models in which the infalling gas has a constant angular velocity at computational boundaries. In these models, the gas envelope has a range of specific angular momenta $0 \le j \le j_\mathrm{inf}$. Figure~\ref{rbinAcc_p02j12c01v10omg2_ug18055904.1.pdf} shows the surface density distribution for the models with $j_\mathrm{inf} = 1.2$ (see Section~\ref{sec:models}). The surface density distribution resembles that for the model with the constant specific angular momentum (the upper right panel of Figure~\ref{rbinAcc_p02j12c01v10_arrow.pdf}) showing asymmetry in the circumbinary disk. The angular velocity of the asymmetric pattern $\Omega_p$ has almost the same value as in the fiducial model (see Table~\ref{table:model-parameters}). 

\begin{figure}
 \epsscale{1.1}
 \plotone{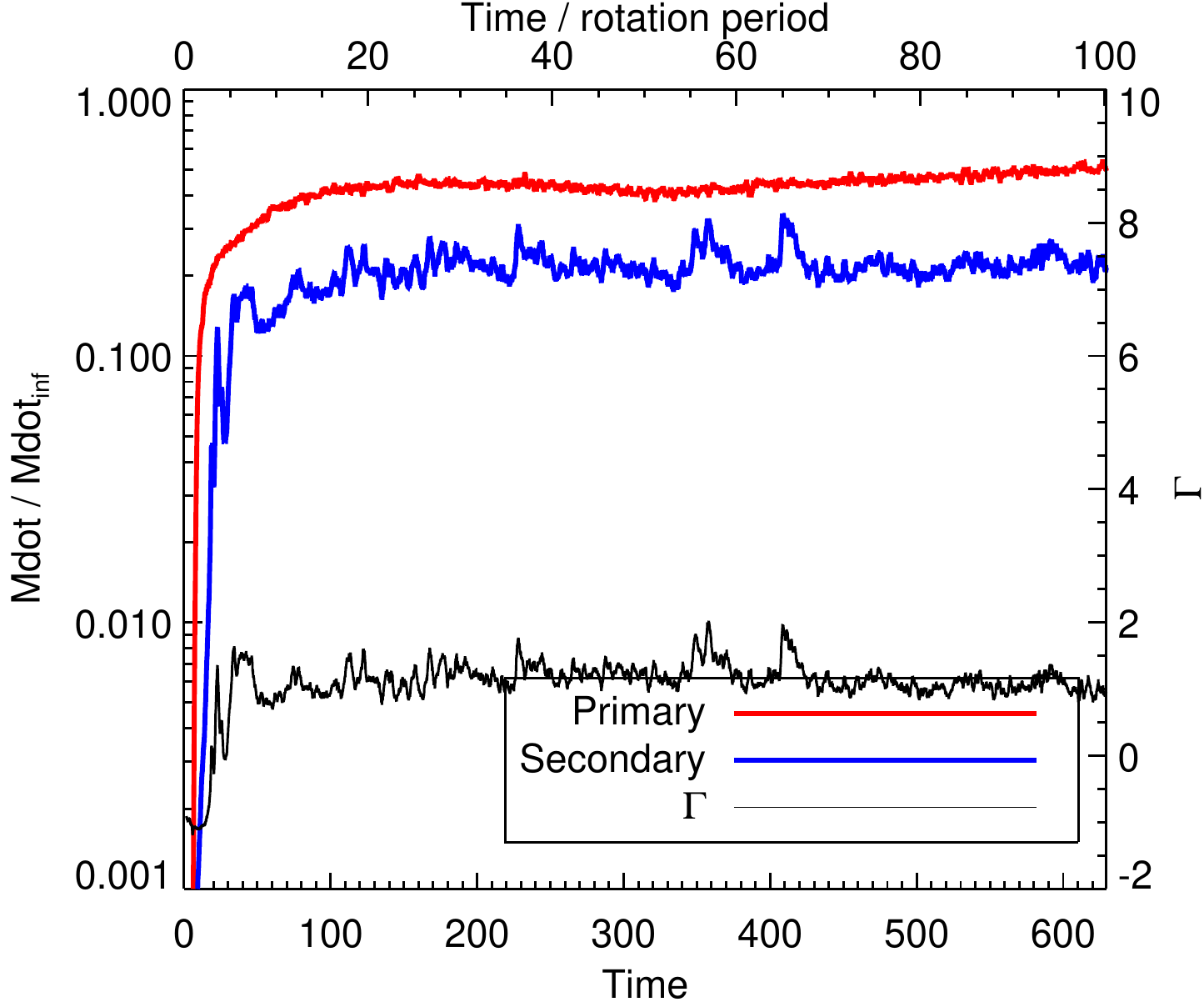}
 \caption{ Accretion rates for the primary (red line) and secondary (blue line) as a function of time for the models with the rigidly rotating envelope model of $(q, j_\mathrm{inf}, c)= (0.2, 1.2, 0.1)$. The accretion rates are normalized by the mass injection rate of the boundary condition. The time is normalized by the rotation period of the binary. The change rate of the mass ratio $\Gamma$ is also shown (black line).
 \label{plot_pmass_p02j12c01v10omg2.pdf}
 }
\end{figure}

The spiral arms remain in this model, but their density contrast is lower than that in the fiducial model because the gap has a higher density. This is because gas with low angular momentum falls along the polar directions and fills the gap. Such gas increases the accretion rate for the primary. The primary has an accretion rate that is two times higher than the secondary (Figure~\ref{plot_pmass_p02j12c01v10omg2.pdf}). The change rate for the mass ratio is low, $\Gamma \sim 1$, but it is still positive. 

For the envelope model with a higher rotation of $j_\mathrm{inf} = 2.0$, the circumbinary disk evolves similar to that for the corresponding model with a constant specific angular momentum (the lower-right panel of Figure~\ref{rbinAcc_p02jxxc01v10_ug.pdf}) in the early stages. After $\sim 60$ revolutions, significant asymmetry of the $m=1$ mode appears at $R \sim 2-3$. As in the previous model, the gap has higher density than in the model with the constant specific angular momentum. The primary star has a higher accretion rate than the secondary star by a factor of $2-3$ throughout the evolution. These differences mainly come from gas with low angular momentum falling from the polar regions.

The three models in this subsection demonstrate that the accretion rates for the binary stars are sensitive to the angular momentum of the infalling envelope.

\section{Discussion}
\label{sec:discussion}

\subsection{Origin of Asymmetry in Circumbinary Disks}
\label{sec:discussion_asymmetry}

As shown in the fiducial model (Figure~\ref{rbinAcc_p02j12c01v10_ug7215464_sigma.2.1.pdf}), the density bump and vortex coexist in the same location. Moreover, a ridge of high specific angular momentum exists at the same angular position relative to the origin. This configuration can be explained as follows. The gravitational torque of the binary acts on the gas close to the inner edge of the gap to increase its specific angular momentum. The gas with enhanced angular momentum moves outward radially, exchanging the radial positions of the gas. This motion is observed as the vortex around the density bump. 

This is similar to the situation where Rayleigh's criterion \citep[e.g.,][]{Chandrasekhar61} is violated; a disk with a negative gradient of specific angular momentum is highly unstable. Since the initial state with a negative gradient of specific angular momentum is unstable, such a state may not appear in the numerical simulations. 
The violation of Rayleigh's criterion is sufficient to cause Rossby wave instability (RWI) in many cases, which causes  vortices and pressure bumps \citep{Li0,Ono16,Ono18}. The RWI requires the vortensity, $\left(\nabla \times \bm{v}\right)_z/\Sigma$, to have a local minimum for a barotropic gas \citep{Lovelace99,Ono16}. We confirmed that the vortensity in our models has a local minimum at the location of the vortex.

The asymmetry does not result from keeping the angular momentum of the infalling gas constant. During several rotations of the binary, the gravitational torque produces an angular momentum distribution in the circumbinary disk. Consequently, the disks extend above the centrifugal radius for the original angular momentum, $R_\mathrm{cent} = j_\mathrm{inf}^2/(GM_\mathrm{tot})$. Note that the models with the rigidly rotating envelope exhibit asymmetry in the circumbinary disk. In these models, the infalling gas has a range of angular momenta. 

\subsection{Ubiquity of Asymmetry in Circumbinary Disks}

For a binary system with a circumbinary disk, the stars rotate with higher angular velocity than that of the circumbinary disk. Therefore, the circumbinary disks receive angular momentum from the stars via gravitational torque. The spiral arms, gaps, and asymmetry in the circumbinary disks are natural outcomes of binary systems with circumbinary disks. We also note that a circumbinary disk has a higher angular momentum than the infalling envelope as a result of such gravitational torque (e.g., Figure~\ref{rbinAcc_p02j12c01v10_ug7215464_sigma.2.1.pdf}).

In our models with very high specific angular momenta ($j_\mathrm{inf} \gtrsim 1.8$), the centrifugal radii are large enough for the gravitational torque to act weakly on the circumbinary disks. When we consider the rigidly rotating envelope model, which may be sometimes realistic, the asymmetry appears as described in Section~\ref{sec:effect_of_infalling_envelope}. Moreover, when we consider a binary orbit with high eccentricity, the binary may interact with the circumbinary disk effectively, even in high angular momentum cases \citep[e.g.,][]{Dunhill15,Price18}. This results in asymmetry in the circumbinary disk. 

\subsection{Application to Observations}

Recent high-resolution observations have revealed circumbinary disks with asymmetric structures. \citet{Takakuwa17} observed the asymmetric circumbinary disk around the protobinary L1551~NE and compared their observations with numerical simulations based on the models described in this paper. \citet{Takakuwa14} analyzed the density and velocity structures in the circumbinary disk and found evidence that they were influenced by gravitational torque. Such gravitational torque is a driving force behind the asymmetry of the circumbinary disk, as shown in Section~\ref{sec:vortex}.

The young binary UY~Aur exhibits asymmetric brightness in the circumbinary disk, as seen in near-infrared scattered light \citep{Hioki07}. \citet{Hioki07} examined a model where the near side of the disk is brighter than the far side because of forward scattering by the dust. The model requires a thick circumbinary disk with a disk opening angle of $\Phi_\mathrm{open} = 45 \degr$. Our 3D models reproduce thick circumbinary disks as shown in the lower panels of Figure~\ref{show_amr_grid.pdf} and the right panel of Figure~\ref{rbinAcc_p02j12c01v10_ub7215464.pdf}, which are likely to be satisfied by the condition of the opening angle, depending on the disk surface for scattering light. The opening angle becomes small when light is scattered at high density. The large thickness of such circumbinary disks is attributed to the spiral arms, which disturb the disk and increase the vertical velocity of the disks. 

The young binary HD~142527 has an asymmetric circumbinary disk showing a horseshoe dust structure \citep[e.g.,][]{Fukagawa13}. Not only dust, but also gas, seems to have an asymmetric structure with a significant contrast level \citep{Boehler17}. The gas contrast has been reported as 3.75 and is consistent with the contrast of the fiducial model, as shown in Section~\ref{sec:asymmetry_in_cbd}. However, \citet{Lacour16} indicated that the companion is unlikely to be responsible for the gap of the circumbinary disk because of the small apocenter distance compared to the gap radius. The disk-companion interaction is likely to be too weak to induce asymmetry in the circumbinary disk via the mechanism described in this paper. Such a weak interaction is seen in the models with a high specific angular momentum (see lower-right panel of Figure~\ref{rbinAcc_p02jxxc01v10_ug.pdf}), where the asymmetry is unclear. Recently, \citet{Price18} reproduced the asymmetric circumbinary disk by assuming that the companion with highly eccentric orbits interacts with the circumbinary disk.

Recent observations have resolved each of the circumstellar disks in young binary systems \citep[e.g.,][]{Lim16a,Lim16b,Takakuwa17}. The masses of circumstellar disks are sensitive to the gas supply from the infalling envelopes as well as the circumbinary disks. When the infalling envelope consists of gas with a large angular momentum, the circumsecondary disk tends to have a larger mass than the circumprimary disk. In contrast, when the infalling envelope has a low angular momentum even in part, it contributes to the growth of the circumprimary disk because the gas falls onto the central region of the binary. The disk masses are also affected by the viscosity of the disk, which is not taken into consideration here. 

All the circumstellar disks in the simulations have clear outer edges. Such circumstellar disks with sharp outer edges have been observed in the young binaries L1551~IRS5 and L1551~NE by VLA observations \citep{Lim16a,Lim16b}. The simulations indicate that the sizes are insensitive to the gas temperature and the angular momenta of the infalling envelopes. The sizes of the circumstellar disks reflect the those of the Roche lobes. In other words, the tidal truncation controls the disk sizes, as suggested by \citet{Lim16a,Lim16b}.

\subsection{Implications for Planet Formation}

Circumbinary planets are considered to be formed in circumbinary disks. The vortices in stellar disks can capture solid particles and initiate the formation of planetesimals \citep{Barge95,Klahr97}. 
Vortices and density and pressure bumps exist in our simulations. 
The vortex persists in the circumbinary disk, as seen in the simulations here, while it tends to migrate inward in the circumstellar disk around a single star \citep{Richard13}. This is because the vortex shown here is continuously excited by the gravitational torque of the binary.

Turbulence hinders the formation of planetesimals. The turbulent feature in the circumbinary disk calms down after accretion onto the circumbinary disk is terminated (Figure~\ref{rbinAcc_p02j12c01v10-00_ug24427505.1.pdf}). After that, the vortex in the circumbinary disk still exists. This suggests that the formation of planetesimals proceeds more efficiently after the main accretion phase. 

\subsection{Comparison with Previous Studies}

\citet{Shi12} also observed asymmetry in the circumbinary disks of an equal mass binary using MHD simulations. They assumed a relatively cold gas with $c=0.05$ and a weak magnetic field of $\beta = 100$. According to their simulations, the angular velocity of the asymmetry is $\Omega_p \simeq 3.2\times 10^{-3} \Omega_\star$, which is significantly smaller than those of our models with moderate temperatures ($c \ge 0.05$), but consistent with our cold models ($c \le 0.02$). \citet{Shi15} doubled the sound speed ($c=0.1$) and still observed asymmetry in the circumbinary disk. The angular velocity of the asymmetry had higher values of $\Omega_p = (0.17 - 0.21)\Omega_\star$, which are roughly consistent with our moderate temperature models. The difference probably comes from the fact that the gap in the circumbinary disk was located outside the computational domain in \citet{Shi15}, and the gas flow inside the gap was not reproduced. 

\citet{Ochi05} investigated accretion rates onto binary stars by using 2D numerical simulations. They argued that the primary accretes more than the secondary even when more gas falls through the L2 point than through the L3 point. This means that the gas was transferred from the secondary to the primary through the bridge between the primary and secondary. 
In contrast, our simulations show that the secondary has a higher accretion rate than the primary in almost all the models examined here, and only the model with very low angular momentum $j_\mathrm{inf} = 0.5$ shows a negative value of $\Gamma$ (Table~\ref{table:model-parameters}). Moreover, the gas tends to flow from the primary to the secondary through the bridge, as shown in Figure~\ref{rbinAcc_p02j12c01v10_ua7215464_rotframe.pdf} (right) and Figure~\ref{rbinAcc_p02j12c01v10_ua7215464_vp.pdf} (right). The results for the accretion rates in this paper are consistent with \citet{Bate97} and \citet{Young15a}.

\citet{Satsuka17} recently performed 3D simulations to investigate accretion onto the seeds of binary stars. In their models, an infalling envelope changes its density and angular momentum based on a collapsing cloud core model as time proceeds. Because they calculated the models for the early phase ($t \lesssim 20 - 50$), some models exhibited spiral arms without forming  circumbinary disks. Their results may reflect the initial conditions of short-term evolution.

\subsection{Simplification of the Models}
\label{sec:discussion_modellimit}

The models examined here ignore the self-gravity of the gas. The self-gravity is important just after fragmentation of collapsing cloud cores, where the gas envelope around the fragments is more massive than the fragments. The orbits of the fragments change as time proceeds because of the interactions between the fragments and the gas envelope and between the fragments themselves, as shown by \citet{Matsumoto03} and \citet{Matsumoto15b}, who performed numerical simulations while taking self-gravity into account. In order to reproduce the evolution of binaries in the early phase, not only self-gravity but also its back-reaction on the orbits of the stars (or fragments) should be taken into consideration. 

  In the case of the protobinary L1551~NE, the mass of the circumbinary disk is $0.009 - 0.076 M_\sun$, and the masses of primary and secondary stars are $0.68 M_\sun$ and $0.13 M_\sun$, respectively \citep{Takakuwa17}. The disk-star mass ratio is therefore  $M_\mathrm{CBD} / (M_1+M_2) = 0.01 - 0.09$, indicating that the circumbinary disk is not likely to be gravitationally unstable \citep{Kratter16}. However, the mass of the circumbinary disk could be comparable to that of the secondary star when it has the upper limit value. In this case, the orbital motion of the secondary star may be affected by the gravity of the circumbinary disk. 

We assume the isothermal equation of state, ignoring temperature distribution in the gas. The temperature distribution of the gas should be determined by many factors: compression heating due to infalling gas and shock waves, radiation heating from protostars, cosmic ray heating, and cooling by dust and molecular emission. Modeling the realistic temperature distribution increases a number of model parameters, and it also increases computational costs for simulations on long-term evolution. The dynamics of the circumbinary gas is significantly affected by the gas temperature as shown in Figure~\ref{rbinAcc_p02j12cxxv10_ug.pdf}. The isothermal assumption is useful to investigate the dependence on the temperature.

We examined the models in which the infall gas is in the steady state. This results in an infalling envelope with a spatially constant specific angular momentum. Such infalling envelopes have been suggested by observations of the envelopes around protostars \citep[e.g.,][]{Ohashi97}. 

The observations have also suggested that the infalling envelopes increase their specific angular momenta during their evolution \citep{Yen11,Yen13,Yen17}. The models examined here do not include time-dependent infall into the system. When considering a binary with a separation of less than $\sim 100$~au and masses of $\sim 1 M_\sun$, the rotation period is less than $10^3$~years, which is considerably shorter than the period of the accretion phase, i.e., $10^{4-5}$~years, thereby indicating that our quasi-steady state models are applicable to such binary systems. 

The magnetic fields redistribute the angular momenta in the circumbinary disk and circumstellar disks. If we take the magnetic fields into account, the accretion rates for both the stars will increase. The extent of the circumbinary disks can be influenced by the magnetic fields. Change in the angular momentum distribution can affect the orbital evolution of binaries \citep[e.g.,][]{Zhao13}. The effects of the magnetic fields depend highly on magnetic diffusion, which is controlled by the degree of ionization of the gas. The models examined here correspond to the case where the degree of ionization is low. The effect of the MHD will be investigated in future work.

\section{Summary}
\label{sec:summary}

We investigated the structure of a circumbinary system, including the circumbinary disks and circumstellar disks by using hydrodynamical simulations. The fixed mesh refinement method was adopted in order to obtain high resolution around the binary stars. 

\begin{enumerate}
\item A circumbinary disk is formed in the case where infalling gas has an angular momentum higher than the critical value given in Equation~(\ref{eq:cbd_condition}). The spiral arms are excited by association with the circumstellar disks of the primary and secondary. A bridge structure is formed between the circumstellar disks because of the shock wave associated with converging flows.
\item The flow inside the gap is similar to the tadpole orbit in the Roche potential. The gas infalls along the spiral arms in the gap, while gas outflows along them in the circumbinary disk because of the higher angular momentum produced by the gravitational torque of the stars.
\item An asymmetric structure appears in the circumbinary disk. The asymmetric structure rotates with an angular velocity $\Omega_p \sim \Omega_\star / 4$, indicating a corotation resonance with the binary orbit. The angular velocity $\Omega_p$ is independent of the mass ratio $q$ and the specific angular momentum of the infalling gas $j_\mathrm{inf}$ for models with a moderate temperature of $0.05 \le c \le 0.175$. In the cold models with $c \le 0.02$, a narrow eccentric circumbinary disks forms, and the precession rate is approximated by the secular motion of ballistic particles. The hot model with $c \ge 0.2$ exhibits an extended circumbinary disk and the asymmetry rotates more slowly than in the case of the moderate temperature models.
\item The surface density contrast for the asymmetry is weaker in models with a higher temperature. It is also weak in models with a high angular momentum of infalling gas ($j_\mathrm{inf} \gtrsim 1.8$). It is almost independent of the mass ratio $q$.
\item The asymmetric structure is attributed to the gravitational torque of the stars. The gravitational torque increases the specific angular momentum of the circumbinary disk, producing a vortex and density bump there. Due to the gravitational torque, the circumbinary disk has a higher specific angular momentum than the infalling envelope.
\item The disk masses and the accretion rates onto the binary stars are affected by the gas accretion from the envelope into the binary system. In particular, the angular momentum of the envelope controls the accretion rates onto the binary stars. This is consistent with the results reported by many authors \citep[e.g.,][]{Bate97,Young15a}. The angular momentum distribution of the envelope also influences the accretion rates. Gas with low angular momentum contributes to the accretion rate of the primary even if it is partially included in the infalling envelope.

\end{enumerate}

\acknowledgments 
We would like to thank J. M. Stone, T. Hanawa, S. Inutsuka, T. Tsuribe, and T. Ono for fruitful discussions.
Numerical computations were carried out on a Cray XC30 (ATERUI) and XC50 (ATERUI II) at the Center for Computational Astrophysics (CfCA), National Astronomical Observatory of Japan (NAOJ), and on a FUJITSU FX100 (HOKUSAI-GreatWave) at the Advanced Center for Computing and Communication (ACCC), RIKEN.
This research was supported by the Japan Society for the Promotion of Science (JSPS) KAKENHI Grant Numbers 18H05437, 18K03703, 17K05394, 17H02863, 16H07086, 26400233, and 26287030, and by the NAOJ Atacama Large Millimeter/submillimeter Array (ALMA) Scientific Research Grant Number 2017-04A.

\end{document}